\documentclass{aa}
\usepackage{graphicx}
\usepackage{txfonts}
\usepackage[dvipsnames]{xcolor}
\usepackage{amssymb}
\usepackage{siunitx}
\usepackage{soul}
\usepackage{ulem}
\usepackage[hidelinks]{hyperref}
\hypersetup{
    colorlinks=true,
    linkcolor=blue,
    filecolor=blue,      
    urlcolor=blue,
    citecolor=blue,
    }
\usepackage{multirow}

\defcitealias{Zhu_2025}{ZP25}
\defcitealias{Johnstone_2015}{JG15}
\defcitealias{Binder_2026}{Binder et al. 2026}

\begin{document}
\title{The X-ray catalogue of FGK stars within 10\,pc}
\subtitle{The coronal temperature-brightness relation explained with the RTV scaling law}

\author{M. M. Bennedik\inst{1} \and B. Stelzer\inst{1} \and M. Caramazza\inst{1} \and S. Orlando\inst{2} \and A. Binks\inst{1} \and J. Robrade\inst{3}\fnmsep\thanks{In Memoriam: Our dear colleague, Jan Robrade.}}

\institute{$^1$Institut für Astronomie und Astrophysik, Eberhard Karls Universität Tübingen, Sand 1, 72076 Tübingen, Germany\\
\email{bennedik@astro.uni-tuebingen.de}\\
$^2$ INAF -- Osservatorio Astronomico di Palermo ``G.S. Vaiana'', Piazza del Parlamento 1, 90134 Palermo, Italy\\
$^3$ Hamburger Sternwarte, University of Hamburg, Gojenbergsweg 112, 21029 Hamburg, Germany
}
  
\date{Received ; accepted }

\abstract{A comprehensive measurement of stellar X-ray emission has important implications for our understanding of stellar dynamos, exoplanet atmosphere loss, and evaporation of protoplanetary disks. However, the samples discussed in the literature are often biased and the X-ray properties reported in published catalogues are often not homogeneously derived. We present a volume-limited catalogue of X-ray detections of the FGK-type main-sequence stars within 10\,pc, which properly accounts for stellar multiplicity and considers the instrumental effect of optical loading of the X-ray detectors. The FGK\,10pc sample derives from a recent {\it Gaia}-based catalogue of all stars in a sky volume of 10\,pc around the Sun from which we extract all stars of spectral type F, G and K.
To build the X-ray catalogue, we retrieved and analysed all observations from SRG (Spectrum Roentgen Gamma)/eROSITA and \textit{XMM-Newton} for stars in the FGK\,10pc sample, and we cross-matched the sample with ROSAT catalogues. After accounting for optical loading, flares, and multiplicity, we fit thermal plasma models to the X-ray spectra to derive X-ray fluxes and coronal temperatures. The FGK\,10pc sample comprises $60$ stars and has an X-ray completeness of $95$\%. We present a detailed investigation of the relation between coronal temperature and X-ray surface flux and we find the result to be consistent with the ranges covered by different types of coronal magnetic structures observed on the Sun. The lower end of the temperature-brightness relation is defined by the Maunder minimum star HD\,166620 and three other stars whose positions identify them as possible Maunder minimum candidates. We identify systematic differences between the temperature-brightness law derived from \textit{XMM-Newton} and eROSITA data that we attribute to their different instrumental response functions. For the first time, we study the impact of variability, both short-term flaring and long-term activity cycles, on the evolution of coronal temperature along with X-ray brightness. With the exception of three stars, a remarkably small scatter is observed in the temperature-brightness relation across the whole FGK\,10pc sample. A natural explanation of the empirical $kT$-$F_{\!\rm X}$ relation is provided by the Rosner-Tucker-Vaiana (RTV) scaling laws defined for solar hydrostatic magnetic loops. The small spread of the observed relation indicates a universal scaling between the length and filling factor of coronal loops. Three observed outliers to the relation are stars which deviate from this invariance of the ratio of loop length to filling factor for a yet unknown reason.
}

\keywords{X-rays: stars, stars: activity, stars: coronae, stars: magnetic field}
\maketitle

\section{Introduction}\label{sect:introduction}
The X-ray emission from late-type stars has been recognised to be a key diagnostic of the stellar dynamo in analogy to our Sun \citep{Rosner_1985}. More recently, stellar coronal X-rays have also emerged as an important driver of photoevaporation in protoplanetary disks \citep[see e.g.][]{Owen_2012, Ercolano_2021} and of exoplanet atmospheric escape \citep[see e.g.][]{Lammer_2003, SanzForcada_2011}. As a result of the dynamo's dependence on stellar interior structure \citep{Pallavicini_1981, Brun_2017}, coronal X-ray properties depend on the stellar mass \citep[see e.g.][]{Preibisch_2005}. 

Statistical studies of coronal X-ray characteristics \citep[see e.g.][]{Vaiana_1981, Pizzolato_2002, Wright_2011} as well as studies that evaluate the impact of stellar X-ray emission on disks \citep[e.g.][]{Monsch_2018} and planets \citep[e.g.][]{Foster_2022} usually rely on flux-limited X-ray observations. By definition, flux-limited samples are affected by the Malmquist bias, where brighter stars are visible at greater distances and thus over-represented. Numerous case studies on the X-ray properties of individual known exoplanet systems have been published \citep[see e.g.][for recent examples]{Poppenhaeger_2020, Pillitteri_2022, Maggio_2023, Acharya_2023}. Clearly, this approach is unfeasible for every single planet system that gets discovered. Assessing the full range of X-ray activity levels spanned by stars of a given spectral type (SpT) or mass is, therefore, critical to understand the mass-dependent evolution of stellar spin, which rules the dynamo, and to identify the conditions for planet formation and evolution. This can be achieved only by volume-limited samples. Since the sky volume accessible to sensitive enough X-ray observations is small, general conclusions rely on the assumption that such nearby samples are representative for the broader environment of our Galaxy. With this assumption, the X-ray properties of a volume-limited sample can constrain the range of conditions under which disks and planets evolve \citep[e.g.][]{Zhu_2025, Zheng_2026}.

During the ROSAT era, the NEXXUS data base by \citet{Schmitt_2004} contained the first volume-limited X-ray samples of stars of SpT F and G up to distances of ${d=14\,\mathrm{pc}}$, and K- and M-type stars up to distances of ${d=12\,\mathrm{pc}}$ and ${d=6\,\mathrm{pc}}$, respectively. With newer facilities like \textit{XMM-Newton} and eROSITA, more sensitive X-ray data has become available. Using data from these missions combined with the {\it Gaia} census of the solar neighbourhood, \citet{Caramazza_2023} presented the X-ray properties of the volume-limited sample of M dwarfs within $10\,\mathrm{pc}$, with the faintest X-ray emitters shown to share properties with solar coronal holes. Expanding this sample with \textit{Chandra} data, \citet{Zhu_2025} (henceforth \citetalias{Zhu_2025}) defined samples containing the ``best'' X-ray measurements for G stars (${d=20\,\mathrm{pc}}$), K stars (${d=16\,\mathrm{pc}}$), and M stars (${d=10\,\mathrm{pc}}$).

In \cite{Bennedik_2026}, we have defined a sample of all FGK-type stars within $10\,\mathrm{pc}$ as a comparison sample for our X-ray study of the Maunder Minimum star HD\,166620. In this article, we present a comprehensive multi-mission catalogue of X-ray measurements and ancillary stellar parameters for the FGK\,10pc sample. 
As database we use ROSAT, eROSITA, and \textit{XMM-Newton} observations and include a systematic assessment of unresolved binary stars and optical loading of the X-ray detectors. Our scientific focus is on the coronal temperature-brightness relation. Quantifying the connection between these two parameters serves to estimate the plasma temperature for more distant emitters for which X-ray observations yield too few counts for a spectral analysis. Thus, it constitutes an important piece in the assessment of the impact of X-ray irradiation on exoplanets and protoplanetary disks.

An apparently universal power law scaling relation between X-ray temperature and X-ray surface flux was established by \citet{Johnstone_2015} (henceforth \citetalias{Johnstone_2015}) on a sample of $24$ nearby stars of SpT F to M based on high-resolution X-ray spectral data from {\it XMM-Newton} and {\it Chandra}. However, from a study of nearby M dwarfs detected in the eROSITA eFEDS field, \cite{Magaudda_2022} found a systematic deviation from the \citetalias{Johnstone_2015} law. While the overall correlation between coronal brightness and temperature appears solid, the suspicion is that different instrumental properties and analysis methods may influence the derived power law parameters. Since the majority of stellar X-ray observations are carried out with low-resolution instruments, an instrument-specific calibration of the relation is essential for applications on X-ray faint and distant stars. Here, we construct this empirical X-ray temperature-brightness relation based on CCD-resolution X-ray spectra from {\it XMM-Newton}'s EPIC/pn and eROSITA from the FGK\,10pc sample and compare the result to that obtained by \citetalias{Johnstone_2015}.

Observations of stellar coronae are unresolved. Due to this limitation, our understanding of stellar corona rely on the Sun, the only corona that can be resolved in detail. In this context, the solar measurements presented by \citetalias{Johnstone_2015} are particularly relevant, as they showed that the solar minimum and maximum of the solar 11-year cycle are roughly consistent with their stellar temperature-brightness power law relation. We extend this solar-stellar comparison to the X-ray properties of individual magnetic structures in the solar corona. Previous studies \citep[see e.g.][]{Caramazza_2023, Joseph_2026} demonstrated that solar coronal structures, which are classified on the basis of their intensity in X-ray images \cite{Orlando_2001}, can be used to characterise the coronae of other stars. Comparing the typical plasma temperature and surface flux of such structures on the Sun to the same parameters from surface-averaged measurements for other stars, thus, allows us to estimate which type of magnetic regions are present in their coronae.

It is widely accepted that coronal magnetic structures are fundamentally composed of coronal loops \citep[e.g][]{Reale_2014}. \citet{Rosner_1978} introduced the Rosner-Tucker-Vaiana (RTV) scaling laws to describe the properties governing quasi-static coronal loops. Under the assumption of a balance between volumetric heating, thermal conduction, and radiative losses, one of the RTV scaling laws relate coronal loop length, temperature, and pressure. In this article we show for the first time that the empirically observed slopes in the coronal temperature-brightness relation naturally arise from the RTV scaling laws.
 
Our article is structured as follows. In Sect.~\ref{sect:Sample} we define the FGK\,10pc sample. The X-ray database, data extraction and spectrum and light curve analysis are described in Sect.~\ref{sect:Data}. In Sect.~\ref{sect:EmpiricalRelation}, we present the empirical temperature-brightness relation, its instrument-specific power law fits derived from our sample, and compare it to coronal magnetic structures observed on the Sun. Furthermore, we identify possible Maunder minimum candidates and discuss the evolution of coronal flux and temperature throughout activity cycles. In Sect.~\ref{sect:RTV}, we show that the temperature-brightness relation derives from the RTV scaling laws. We summarise our findings in Sect.~\ref{sect:summary}.

\section{Sample}\label{sect:Sample}
\subsection{Sample definition}\label{subsect:sample}

In this study, we build upon the $10$\,pc catalogue which was published by \citet{Reyle_2021} and updated by \citet{Reyle_2023}. The subsample we study in this work has been defined by \citet{Bennedik_2026}. In brief, we selected all stars of SpT F, G, and K according to the SpT given in \citet{Reyle_2021}. To achieve completeness when combining our sample with the M10\textsc{pc-gaia sample} sample defined by \citet{Caramazza_2023}, we additionally include the M0 dwarf HD\,232979 as it is a K9 dwarf according to its \textit{Gaia} ${G_{\rm{BP}}-G_{\rm{RP}}}$ in the table maintained by E. Mamajek\footnote{See\,\,\url{https://www.pas.rochester.edu/~emamajek/EEM_dwarf_UBVIJHK_colors_Teff.txt}} and is, therefore, not considered in the M10\textsc{pc-gaia sample}. Due to the proximity of the sample and the sensitivity of \textit{Gaia}, no stars are expected to be missing from the $10$\,pc catalogue by \citet{Reyle_2021}. Four stars of the sample selected as described above are identified as subgiants by \citet{Reyle_2021} and are excluded from our analysis. Furthermore, we exclude $\alpha$\,CMi\,A as a known subgiant \citep[][]{Bond_2015}.

\begin{figure}
    \centering
    \includegraphics[width=0.45\textwidth]{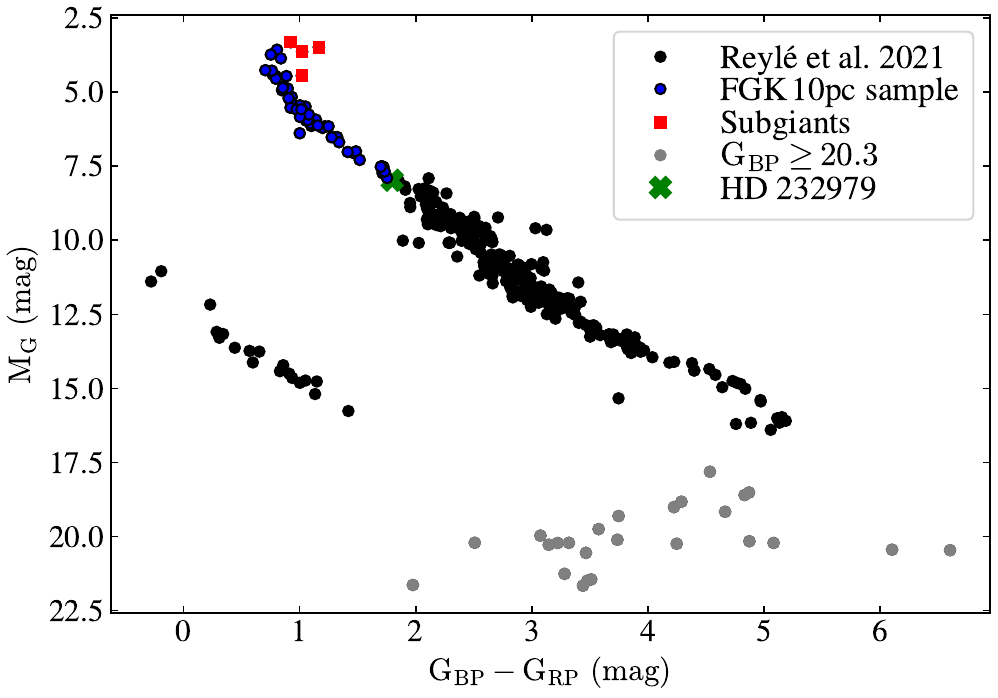}
    \caption{\textit{Gaia} colour-magnitude diagram for the complete 10\,pc catalogue by \citet{Reyle_2021} with the FGK\,10pc sample. Excluded from the sample are the subgiants. The M0-dwarf HD\,232979 is included in the sample. Faint sources with $M_G\lesssim17\,\mathrm{mag}$ have unreliable photometry.}
    \label{fig:Sample}
\end{figure}

A total of $60$ stars are contained in the FGK\,10pc sample, of which $7$ stars do not have photometry in the \textit{Gaia} Data Release 3 \citep{Gaia_2023} either because they are too bright or because they are unresolved companions. In Fig.~\ref{fig:Sample} we present the FGK\,10pc sample in the \textit{Gaia} colour-magnitude diagram displaying for context the full 10\,pc catalogue by \citet{Reyle_2021} for all stars with \textit{Gaia} photometry. L- and T-type stars with $\mathrm{G_{BP}}\!\!\geq\!\!20.3$\,mag have unreliable \textit{Gaia} photometry \citep{Riello_2021} and are marked in grey.

To determine the bolometric luminosities and effective temperatures of stars in the FGK\,10pc sample, we used isochrones from the \textsc{parsec v2.0} database \citep{Nguyen_2022, Nguyen_2025}. We described the procedure in detail in \citet{Bennedik_2026}. In short, we fit each star to a grid of isochrones covering ages of 0.5--9.0\,Gyr within the \textit{Gaia} colour-magnitude diagram. The observed metallicities listed in Table~\ref{tab:FundamentalProperties} are used to constrain the isochrone for each star, leaving the stellar age as a free parameter. The stellar radii are then computed with the Stefan-Boltzmann law.

\subsection{Multiplicity}\label{subsect:multiplicity}

Some of the stars within the FGK\,10pc sample are part of multiple star systems. Identifying these systems is important for the interpretation of the X-ray data, especially since -- depending on the system separation and the resolution of the instrument -- it might not be possible to separate the observed X-ray emission into the contributions of the individual stars. We determine the multiplicity for each system with the updated catalogue by \citet{Reyle_2023}. Comparison with the recent study by \citet{Gonzalez_Payo_2026} shows agreement for all systems in the FGK\,10pc sample, except for HD\,50281, where the close binary low-mass companion of the M2 dwarf HD\,50281\,B is questioned. The A and B components of this system are resolved in the X-ray observations, and as the B component is not part of the sample, this difference is not relevant for our analysis. For the following counts we reject the companion of HD\,50281\,B.

In total, $36$ out of $60$ stars within the FGK\,10pc sample are part of multiple systems. Some companions in these systems are not themselves part of the sample. Accounting for this, the sample contains $20$ stars in $13$ binaries, $12$ stars in $8$ triples, one star in a quadruple system, and three stars in one quintuple system.

\section{X-ray data and analysis}\label{sect:Data}
\subsection{X-ray database}\label{subsect:Xraydata}

In \citet{Bennedik_2026} we had compiled X-ray data for the FGK 10pc sample. This comprised data from the Second ROSAT All-Sky Survey Point Source Catalogue (2RXS) \citep{Boller_2016}, the Second ROSAT Source Catalogue of Pointed Observations (2RXP) \citep{Rosat_2000}, the eROSITA Data Release 1 \citep{Merloni_2024} and the 4XMM-DR14 catalogues \citep{Webb_2020}. Here we extend the X-ray database and its analysis. 
We extracted data from the first five eROSITA all-sky surveys (eRASS1--5), instead of using the products from the eROSITA Data Release 1, which is limited to eRASS1. These are limited to the western Galactic hemisphere for which the German eROSITA consortium has data rights. eRASS5 was discontinued and is therefore only available for a part of the sample. Regarding \textit{XMM-Newton}, we do not use the products of the 4XMM-DR14 catalogue for this study in order to perform a customised data extraction taking account of variability (see Sect.~\ref{subsubsect:LCs}) and to measure plasma temperatures through spectral analysis. To this end, we acquire publicly available data for all EPIC/pn observations covering the FGK\,10pc sample. Additionally, we conducted two \textit{XMM-Newton} observations for HD\,192310 and GJ\,53\,A\&B (ObsID 0982040301 and 0982040301; PI Bennedik). The extraction of data from eROSITA and \textit{XMM-Newton} is detailed below.

\subsubsection{eROSITA}
For FGK\,10pc stars in the galactic hemisphere with German eROSITA data rights located at ${359.9442^\circ > l > 179.9442^\circ}$ \citep{Merloni_2024}, we downloaded the eRASS sky tiles (approximately ${3^\circ\!\times3^\circ\!}$ regions) produced with pipeline configuration 030 from the internal archive of the Max-Planck-Institut für extraterrestrische Physik (MPE) \citep{Brunner_2022}. These contain merged event files of the seven telescopes on board eROSITA with observations from eRASS1--5 and the combined data for the first four and, where applicable, five surveys (eRASS:4 and eRASS:5). To extract the data, we use the eROSITA Science Analysis Software System (eSASS) user version 240410.0.4. We limit the analysis to the 0.2--4.5\,keV energy band. For lower energies, the sensitivity of eROSITA decreases steeply \citep[see][]{Predehl_2021}, and previous studies have shown that stellar coronae have substantially weaker emission at higher energies with the exception of very large flares \citep[see e.g.][]{Magaudda_2022, Stelzer_2022}. In fact, in the eRASS data of our sample we do not detect any photons with energies above 4\,keV above the background. 

We generate exposure maps, detection masks, and background maps with the \texttt{erbackmap} task with a value of 15 for the Gaussian sigma of the smallest smoothing kernel. These are used as input for the \texttt{ermldet} source detection task, which produces a source catalogue. We use a detection likelihood threshold of $5$. Then, we select the X-ray source with the smallest angular separation to the optical \textit{Gaia} coordinates of each sample star. For this, we propagate the optical coordinates to the average epoch of the corresponding eRASS using the \textit{Gaia} proper motion of the star. In some cases, no X-ray source is found within $30^{\prime\prime}$, implying that the star was too faint for a detection. For each successful match of an X-ray source to an FGK 10pc star, we extract spectra with the \texttt{srctool} task with the built-in automatic determination of source and background regions. Flares are difficult to identify in eRASS data due to the low data cadence (see Sect.~\ref{subsubsect:LCs}) and thus we do not extract and analyse light curves for the individual surveys. In addition, many stars in the sample are too faint for spectral analysis of their quiescent emission because excluding time intervals affected by stellar flares would leave too few counts in individual eRASS. Moreover,  GJ\,570\,A and GJ\,166\,A are partially blended with their M type companions in eROSITA images. Therefore, we manually select their extraction regions as described in Appendix~\ref{app:SpecialExtractions}.

\subsubsection{\it XMM-Newton}
For the extraction of \textit{XMM-Newton} data, we use the \textit{XMM-Newton} Science Analysis Software (SAS) version 22.0.0. The analysis is performed on EPIC/pn data in the 0.2--12.0\,keV energy band. The lower bound is set due to unreliable calibration of the instrument. \textit{XMM-Newton} observations can be affected by solar particle flux, which causes high energy events. To identify the corresponding ``bad time intervals'', we inspect the light curves for photon counts with energies $\geq\!10\,\mathrm{keV}$ on the whole detector by eye and filter out time intervals with significantly higher count rates than the quiescent baseline. Additionally, we retain events with pattern~$\leq\!4$ and mask bad pixels (flag~=~0). Then, we run the source detection. We account for proper motion at the time of observation when matching the optical \textit{Gaia} coordinates of the objects with the detected X-ray sources, which all match well within $10^{\prime\prime}$. To extract light curves, spectra, and instrumental response files for a given star and observation, we define a circular source region centred at the position determined by the source detection and a circular background in an adjacent quiet region void of any sources. The radius of both regions and the position of the background are manually set by visual inspection of the images. The radius of the source region ranges from $11^{\prime\prime}$ for sources with nearby system partners up to $50^{\prime\prime}$ for very bright sources with extended point spread functions (PSF). The background region has a minimum radius of $30^{\prime\prime}$ and is chosen with equal size of the source region if it is larger.

\subsubsection{Optical loading}\label{subsect:opticalloading}
Due to the proximity of the stars in the FGK\,10pc sample, some are optically very bright. Large numbers of optical or UV photons can cause erroneous signals in X-ray detectors, which is commonly known as optical loading. Optical loading does not significantly affect ROSAT observations which used a proportional gas counter as the detector. 

Most \textit{XMM-Newton} observations are not significantly affected by optical loading due to the appropriate selection of filters and readout modes. In the FGK\,10pc sample, the majority of the observations used the \textsc{thick} filter, which absorbs the majority of the low-energy photons and thus reduces optical loading. For the five stars in the FGK\,10pc sample which were observed with the \textsc{medium} filter, we examine the X-ray spectra for optical loading. We presume that a power law model describes the contribution of optical loading based on the study on eROSITA data by \citet{Robrade_2026}. We add a power law component to the model fit of the spectra outlined in Sect.~\ref{subsubsect:SpectralAnalysis}. If the power law fit yields better statistics than a fit without a power law component, we define the observation as affected by optical loading. In some cases, the power law component yields comparable fitting statistics to an additional thermal model component. If this thermal component has a low temperature ($kT\lesssim0.1\,\mathrm{keV}$), we consider it likely that it is caused by optical loading.

\begin{figure}
    \centering
    \includegraphics[width=0.49\textwidth]{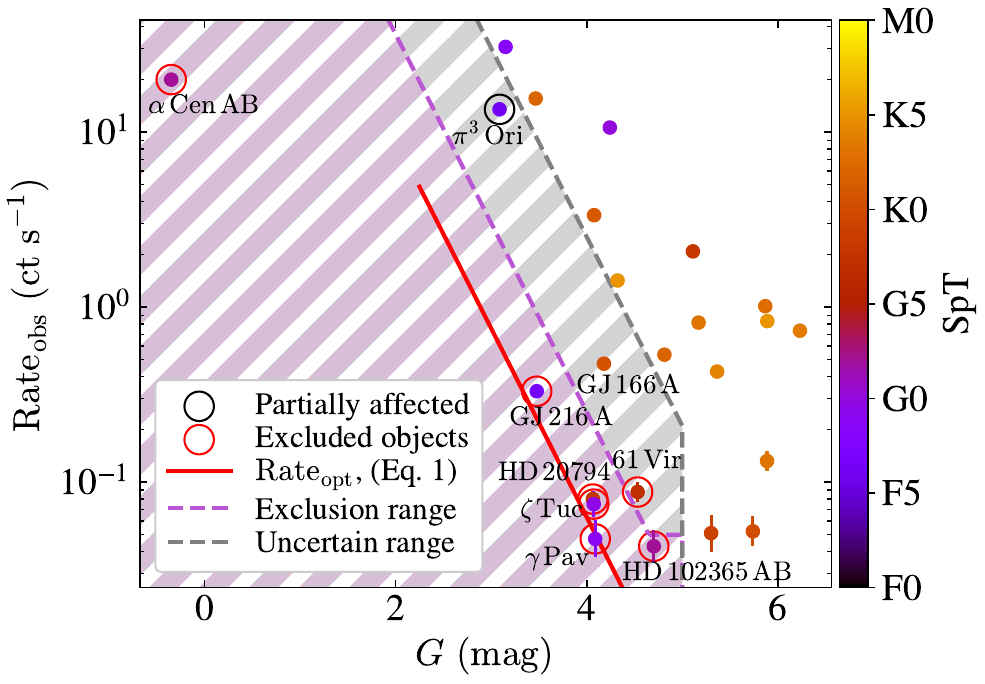} 
    \caption{eRASS:5 count rate versus $G$ magnitude in the FGK\,10pc sample. For unresolved multiple systems, we show the sum of the $G$-band flux and the earliest SpT of the unresolved stars. We discard eRASS data within the purple striped area due to optical loading and examine stars within the grey striped area (see text in Sect.~\ref{subsect:opticalloading} for details). Red and black annuli indicate discarded and partially affected objects.}
    \label{fig:OpticalLoadingCut}
\end{figure}

We find that the \textit{XMM-Newton} observation of HD\,219134 is affected by optical loading, as is presented in Appendix~\ref{app:LoadingSpectra}. With a V-band magnitude of $5.37$, this is the optically brightest among the five stars observed with the \textsc{medium} filter. For the other stars, we do not find evidence of optical loading. We discard the observation of HD\,219134 from further analysis.

eROSITA observations are more commonly affected by optical loading because the whole survey is conducted homogeneously without change of filters. To quantify which eROSITA observations are affected, we use the empirical prediction of the optical contribution to the X-ray count rate, 
\begin{equation}
    \mathrm{Rate_{opt}} = 0.00525 \times 12.0^{5-G}\,\mathrm{(ct\,s^{-1})},
\end{equation}
determined by \citet{Robrade_2026} for sources with a \textit{Gaia} $G$ band magnitude of ${G<5\,\rm{mag}}$. They suggest that observations are significantly contaminated by optical loading if the contribution of $\mathrm{Rate_{opt}}$ is at least one fourth of the total observed count rate $\mathrm{Rate_{obs}}$ or if ${\mathrm{Rate_{obs}} - \mathrm{Rate_{opt}} < 0.05\,\mathrm{ct\,s^{-1}}}$. The analysis of \citet{Robrade_2026} is limited to ${G>2.25\,\rm{mag}}$. For the observations of $\alpha$\,Cen\,A\&B, we extrapolate the relation to lower magnitudes. We present the exclusion range in Fig.~\ref{fig:OpticalLoadingCut}. For multiple stars that are not resolved by eROSITA, we use the sum of the individual $G$ band fluxes and show the earliest SpT that dominates the flux. We conservatively examine all stars for optical loading if they are within one order of magnitude of the cutoff range. We examine whether adding a power law component to the model improves the model fit as described for \textit{XMM-Newton} observations above. The X-ray spectrum of 61\,Vir is consistent with that of an X-ray detection of a star exclusively caused by optical loading as described by \citet{Robrade_2026}. We additionally find that the observations of $\pi^{3}$\,Ori are partially affected by optical loading in the soft regime of its spectra. While we keep these detections in the sample, we exclude them from subsequent analysis. We show spectral model fits for both stars in Appendix~\ref{app:LoadingSpectra}.

In total, $7$ of $24$ X-ray sources detected in eRASS are discarded. All of these sources contain stars with SpT K0 or earlier. On average, spectral model fits (see Sect.~\ref{subsubsect:SpectralAnalysis}) of these source have a coronal temperature of $0.10\,\rm{keV}$. In comparison, sources that were retained in the sample show an average coronal temperature of $0.34\,\rm{keV}$. This suggests that the discarded sources are, indeed, significantly affected by optical loading.

\subsection{X-ray emission from multiple star systems}\label{subsect:XrayMultiples}

For the eROSITA and \textit{XMM-Newton} observations, we determine whether sources in multiple systems as described in Sect.~\ref{subsect:multiplicity} are resolved by inspecting the images and the output of the source detection, whereas for ROSAT observations we adopt a minimum separation of $40^{\prime\prime}$ \citep[see e.g.][]{Neuhaeuser_1995} for two stars to be reliably resolved. For multiple systems that are not resolved in X-ray observations, we treat them based on the SpT of the stars in the system. If multiple system components are expected to be significant X-ray emitters, we treat them as a single source and use a combined identifier (e.g. ``GJ\,66\,AB''). This includes all stars with SpT F--M. We exclude earlier-type stars as no X-ray emission is expected \citep{Stelzer_2006, Schroeder_2007} and later-type stars because of their low X-ray emission caused by high resistivity in their largely neutral photospheres \citep[see e.g.][]{Mohanty_2002}, which would make a negligible contribution to the combined X-ray source. For more details on individual unresolved companions see Appendix~\ref{app:UnresolvedCompanions}.

We find that $24$ stars in $15$ systems of the FGK\,10pc sample are unresolved with all available X-ray instruments. Considering the ROSAT, eROSITA and {\it XMM-Newton} observations, the FGK\,10pc sample contains X-ray detections for $57$ out of $60$ stars, resulting in an X-ray completeness of $95$\%. There are $54$ stars with ROSAT detections, $22$ stars with eROSITA detections, and $19$ stars with \textit{XMM-Newton} detections.

\subsection{Spectral analysis}\label{subsubsect:SpectralAnalysis}

We performed the spectral analysis for observations from \textit{XMM-Newton} and eROSITA with XSPEC\footnote{See: \url{https://heasarc.gsfc.nasa.gov/xanadu/xspec/}} version 12.14.1 interfaced with PyXSPEC \citep{GordonArnaud_2021}. We bin the spectra in various ways, each time with a different number of minimum counts imposed for each bin. The number of minimum counts is varied in the range of 10--40 with a step size of 5 for \textit{XMM-Newton}. For eROSITA, we decrease the lower limit to 5 due to the low number of total counts in some detections. 

We systematically fit each spectrum in the $0.2-10$\,keV energy range with thermal models \citep[\texttt{APEC},][]{Smith_2001} with one, two, and three components and all bin sizes (i.e. 21 fits per spectrum for \textit{XMM-Newton} and 24 fits for eROSITA). Extinction is negligible at the small distances of the stars and thus we set it to zero. We adopt the abundance library of \citet{Asplund_2009} and use a global abundance of $Z=0.3\,Z_\sun$, a value typically found for stellar coronae \citep[see e.g.][]{Robrade_2005, Maggio_2007}. This leaves the coronal plasma temperatures ($kT$) and the emission measures ($EM$) of each spectral component free to be fit, from which we get the best-fit parameters with uncertainties. To obtain the flux in the 0.1--2.4\,keV ROSAT energy band, we convolve the best-fit model with the \texttt{cflux} model component. This is done instead of using the \texttt{flux} routine of XSPEC because it yields more accurate confidence intervals. The emission measure is degenerate with the flux, and therefore we freeze the best-fit emission measure values. Then, we re-run the fit to obtain the uncertainty of the total flux.

For the \textit{XMM-Newton} spectrum of $\beta$\,Com, a model fit with a reduced chi-squared $\chi_{\rm red}^2<3$ was not achieved with any of the \texttt{APEC} models and we used a variable abundance model (\texttt{vapec}) with two temperature components instead. Following \citet{Zheng_2026}, we left the abundances of Fe and the $\alpha$-elements O, Ne, and Si variable while fixing the abundance of other elements to the same $Z=0.3\,Z_\sun$ as above. 

To decide how many thermal components are needed to fit a given spectrum requires quantifying whether there is a statistical improvement when adding an additional component to the model. To this end, starting with the single-temperature model we use the Fisher test \citep{Fisher_1922} to determine whether increasing the number of model parameters leads to a significant improvement of the fit. The Fisher test returns a p-value corresponding to the significance of the improvement. In general, we only add thermal components to the model if there is an improvement with a significance of at least $3\sigma$. Otherwise, the lower-complexity model is kept. A few cases with significances between 3$\sigma$--3.5$\sigma$ have badly constrained model parameters and we adopt the simpler model instead. This procedure is performed separately for the sets of spectra with different bin size.

Finally, we determine which spectral bin size yields the most appropriate model fit. We generally select the binning with a resulting $\chi_{\rm red}^2$ closest to one. In a few cases however, we instead manually select a different binning based on visual inspection of the X-ray spectra or due to more well-constrained model parameters.

For a model with $N$ thermal components $i$, we compute the logarithm of the mean coronal thermal energy of the coronal plasma, $\smash{\overline{kT}}$, by weighting the logarithm of the thermal energy of each component with its $EM$ \citep{Johnstone_2015, Binder_2026}, such that
\begin{equation}\label{eq:kT}
    \mathrm{log}\,\overline{kT} = \frac{\sum_{i=1}^{N}\mathrm{log}\,kT_i \cdot EM_i}{\sum_{i=1}^{N}EM_i}.
\end{equation}
The lower and upper errors are computed as the 16th and 84th percentiles of Monte-Carlo simulations with 10\,000 iterations where each sample is drawn from a two-piece normal distribution of the uncertainties of each $EM_i$ and $kT_i$.

For individual and stacked eRASS detections of 41\,Ara\,A and individual eRASS detections of HD\,100623\,A, the number of net source counts is $<$30 and thus too low for a reliable spectral analysis. For these detections, we use the 0.2--4.5\,keV count rates from the source detection pipeline and multiply them with the conversion factor $CF = 1.06\cdot10^{-12}\,\rm{erg\,cm^{-2}\,cts^{-1}}$ derived by \citet{Bennedik_2026} to obtain the flux in the 0.1--2.4\,keV ROSAT energy band. While the energy band for our count rates is extending to higher energies than the range used by \citet{Bennedik_2026}, the difference is negligible, because the flux with energies above 2.3\,keV in eROSITA spectra contributes less than 1\% even for the hottest coronae in our sample.
We use the distances derived from \textit{Gaia} parallaxes to convert the fluxes to X-ray luminosities, $L_{\mathrm{X}}$.

\subsection{Light curve analysis}\label{subsubsect:LCs}

\begin{figure}
    \centering
    \includegraphics[width=0.45\textwidth]{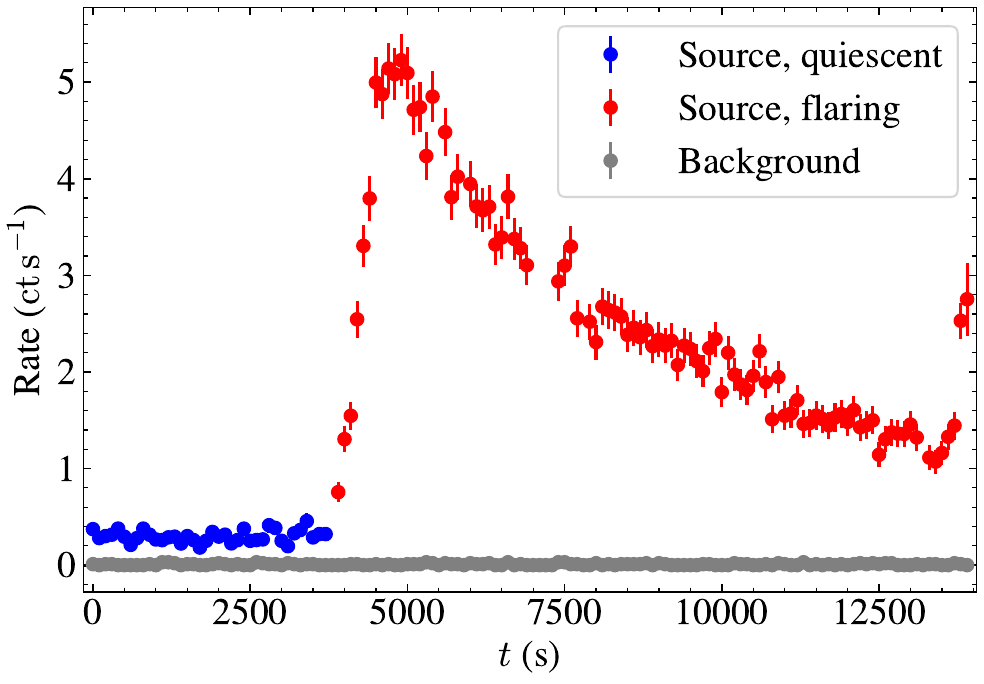} 
    \caption{EPIC/pn light curve of 61\,Cyg\,B as an example for our by-eye separation of quiescent and flaring time intervals. The blue points define the quiescent state and the red points are ascribed to the flaring state.}
\label{fig:LC_Flare}
\end{figure}

For each \textit{XMM-Newton} EPIC/pn observation we extract a light curve with a bin size of $100\,\mathrm{s}$. To identify the quiescent emission, we select time intervals with flares by visual inspection. Then we repeat the data extraction excluding these time intervals, resulting in count rates and spectra only containing quiescent states. Similarly, we extract data containing only the flare intervals. As an example, a light curve of 61\,Cyg\,B is presented in Fig.~\ref{fig:LC_Flare}.

eROSITA light curves have a much lower time resolution compared to \textit{XMM-Newton} due to the scanning strategy during the eRASS. While eROSITA rotates around its axis with a period of ${4\,\mathrm{h}}$ (one ``eRODay''), it observes a given source for up to ${40\,\mathrm{s}}$ per visit \citep{Merloni_2024}. X-ray flares are often shorter than an eRODay. Reliably identifying flares in eRASS light curves is, therefore, not trivial and we leave this task for a future work.

\subsection{X-ray catalogues}\label{subsect:Catalogues}
We present the fundamental properties and mean X-ray parameters of the stars in the FGK\,10pc sample in Table~\ref{tab:FundamentalProperties}. To determine the mean X-ray parameters, we only use the observations in which the components in multiple systems are resolved in the X-ray image. If they are not resolved in any of the available observations, we specify this in the \texttt{X-ray~source} column and normalise the flux with the sum of the surface areas of both stars. From eROSITA we use data from individual eRASS, unless the stacked catalogue (eRASS:5 if the star is included in eRASS5 and eRASS:4 otherwise) is the only available data with sufficient counts for analysis. We only include the quiescent phases during \textit{XMM-Newton} observations as described in Sect.~\ref{subsubsect:LCs}. Only one star from the FGK\,10pc sample, HD\,102365\,A, does not have a measured X-ray flux and an additional two stars do not have reliable X-ray fluxes, that is from observations not affected by optical loading.

Each individual observation, the used spectral model for \textit{XMM-Newton} and eROSITA observations with sufficient counts for spectral analysis, and the derived X-ray parameters are provided in Table~\ref{tab:IndividualObservations}. Here, the X-ray flux and luminosity represent direct results from spectral fitting and are not corrected for multiplicity. All observations that were dropped due to optical loading (see Sect.~\ref{subsect:opticalloading}) are listed in Appendix~\ref{app:NotInSample}.

\subsection{Comparison with \citetalias{Zhu_2025}}

The X-ray properties of late-type stars in the solar neighbourhood have recently been discussed by \citetalias{Zhu_2025} using multi-mission data that strongly overlaps with our data base. However, differences in the sample and analysis distinguish their study from ours that we discuss in the following. 

While \citetalias{Zhu_2025} study GKM-type stars, we include $7$ stars with spectral types F, but we defer the study of M stars to a separate work (Stelzer et al., in prep.). As can be seen from Fig.~4 of \citetalias{Zhu_2025}, our restriction to the 10\,pc volume limits the number of stars with respect to their study. However, the increased sky volume used by \citetalias{Zhu_2025} comes at the expense of completeness as can be seen from their Fig.~3. 

We homogeneously treat the data from all observations. Namely, we have analysed all individual eROSITA and \textit{XMM-Newton} observations of our sample stars, including data from all five eRASS. \citetalias{Zhu_2025} relied on published values with inhomogeneous treatment and they use only a single observation per star choosing the data set with the highest sensitivity and angular resolution or with the longest observation. This impedes an assessment of the role of variability in the overall sample X-ray properties. Contrary to our work they include {\it Chandra} data, but they use only the first of the eRASS.

We systematically identify unresolved multiple systems and assign the observed X-ray emission based on plausibility arguments to the individual components. In contrast, \citetalias{Zhu_2025} exclude such systems. Furthermore, we excluded observations affected by optical loading (see Sect.~\ref{subsect:opticalloading}). \citetalias{Zhu_2025} do not discuss optical loading, which leads to inclusion of wrong values for the X-ray parameters (e.g. for 61\,Vir and HD\,102365). Finally, there are differences in the calculation of the fluxes. For observations without published fluxes, \citetalias{Zhu_2025} rely on published values for plasma temperatures to compute count rate to flux conversion factors from 1T-APEC models with WebPIMMS. If no temperature is published, an assumption for typical values for the SpT is made. As is discussed by \citetalias{Zhu_2025}, this leads to uncertainties for the determined X-ray luminosities. For \textit{XMM-Newton} and eROSITA observations, we avoid uncertainties introduced by conversion factors by measuring the X-ray flux directly with spectral fitting.

\section{Empirical coronal temperature-brightness relation} \label{sect:EmpiricalRelation}

\begin{figure*}
\sidecaption
  \includegraphics[width=12cm]{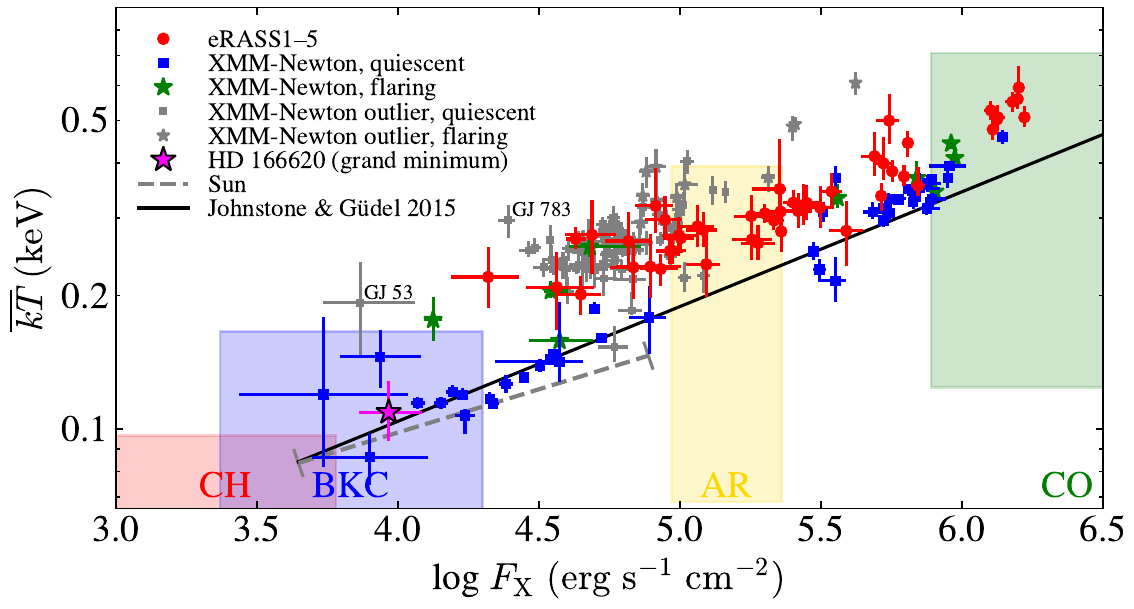}
     \caption{Coronal mean temperature versus X-ray surface flux for all \textit{XMM-Newton} and eROSITA observations. For {\it XMM-Newton} data representing quiescent and flaring states are distinguished. For eRASS the individual surveys are shown. The observations of the stars 61\,Cyg\,A\&B, GJ\,53\,A, and GJ\,783\,A that appear as clear outliers despite not flaring are marked in grey. Typical $F_{\!\rm X}$ and $kT$ ranges for different types of magnetic structures on the Sun are shown as coloured boxes. The range of the Sun during its $11$\,yr cycle adopted from \citet{Peres_2000} is shown as a grey dashed line and the Maunder minimum star HD\,166620 with values determined by \citet{Bennedik_2026} is shown in pink. For details see Sect.~\ref{sect:EmpiricalRelation}.}
     \label{fig:kT-Fx-Main}
\end{figure*}

Our systematic analysis of multi-epoch X-ray spectra for the FGK\,10pc sample enables us to revisit the coronal temperature-brightness relation. As discussed by \citetalias{Johnstone_2015}, for samples spanning a broad range of stellar masses (and hence radii), the X-ray surface flux, $F_{\!\rm X}$, is the best parameter to characterize the intrinsic emission level, because it removes the dependence on stellar size. For the same reason, $F_{\!\rm X}$ enables a direct comparison between stellar and solar activity levels. 

We display the double-logarithmic distribution of $F_{\!\rm X}$ and $\smash{\overline{kT}}$ for the FGK 10\,pc sample in Fig.~\ref{fig:kT-Fx-Main}. Both eROSITA and \textit{XMM-Newton} observations are included, and for each star all individual X-ray detections with a spectrum analysed by us are shown. This means that for {\it XMM-Newton} we consider separately quiescent and flaring states, while for eROSITA our spectral fits average over all activity states within a given eRASS. Besides the FGK\,10pc sample, we show in Fig.~\ref{fig:kT-Fx-Main} the typical ranges measured for the solar corona (see Sect.~\ref{subsubsect:solarregions}) and the only confirmed Maunder Minimum star, HD\,166620 (see discussion in Sect.~\ref{subsubsect:MMstars}). The best-fit relation for the sample of $24$ bright, nearby stars spanning masses from $0.1$ to $1.2\,{\rm M_\odot}$, and hence SpT F to M, presented by \citetalias{Johnstone_2015} is shown in black.

A clear correlation is seen between $\log{F_{\!\rm X}}$ and $\log{\smash{\overline{kT}}}$ for the FGK\,10pc sample. It spans more than two orders of magnitude in surface flux and nearly a factor $10$ in X-ray temperature. Despite the overall well-defined correlation some substructure is seen. First, when only the {\it XMM-Newton} data is considered, a large cluster of outliers upwards/leftwards can be noticed. Almost all of these outliers represent the resolved binary stars 61\,Cyg\,A\&B. We mark these stars as well as the only other quiescent outliers, GJ\,783\,A and GJ\,53\,A, in grey. We also find that the majority of the remaining {\it XMM-Newton} data is dominated by two systems, $\alpha$\,Cen\,AB (blue symbols in the lower left) and $\epsilon$\,Eri (blue symbols in the upper right). Both stars are part of a dedicated discussion of stars with X-ray activity cycles in Sect.~\ref{subsect:XrayVar}. Comparing the {\it XMM-Newton} and eRASS data sets it seems that eRASS defines a relation that is slightly offset to the upper left with respect to EPIC/pn. We discuss this difference in Sect.~\ref{subsubsect:HighLowRes}.

\subsection{Power law fit}\label{subsect:Powerlaw}

\begin{figure}
    \centering  
\includegraphics[width=0.46\textwidth,trim={0 1.69cm 0 0},clip]{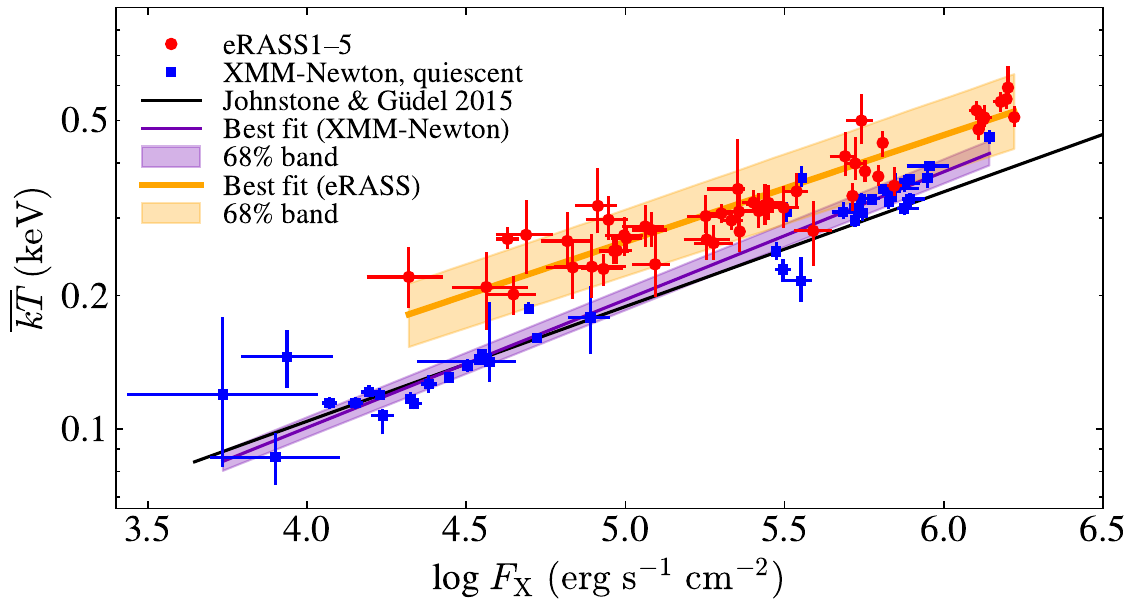} 
 \hfill
\includegraphics[width=0.46\textwidth,trim={0 0 0 0.11cm},clip]{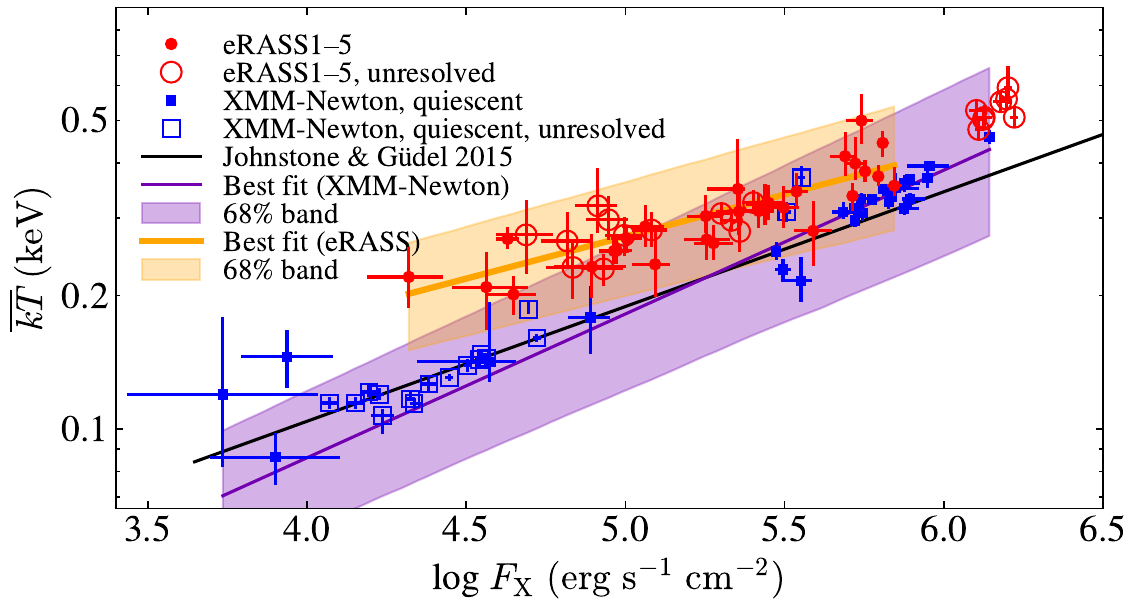}  
    \caption{X-ray temperature-brightness relation excluding systems with peculiar behaviour marked in grey in Fig~\ref{fig:kT-Fx-Main}. \textit{Upper panel:} Fit including unresolved multiple stars. \textit{Lower panel:} Fit excluding unresolved multiple stars (open symbols).}
    \label{fig:kT-Fx-withfit}
\end{figure}

As the data have errors in both dimensions, we fit the temperature-brightness relation with a Monte-Carlo simulation where samples are drawn from a two-piece normal distribution. In each of the 1000 iterations, the sample is fit with an orthogonal distance regression where the weights are determined by the root mean square of the asymmetrical errors. As the best-fit parameters we use the median values and as the uncertainties we use the central 68\% band. In the fit we exclude the flares identified in \textit{XMM-Newton} observations and the three systems marked as outliers in Fig.~\ref{fig:kT-Fx-Main}. We treat {\it XMM-Newton} and eROSITA data separately because of our by-eye suspicion of a different distribution. To enable a direct comparison with \citetalias{Johnstone_2015}, we fit the data in linear space and adopt the relation
\begin{equation}\label{eq:T_Fx}
\smash{\overline{T}}_{\rm cor} = a\cdot F_{\!\rm X}^{b},
\end{equation}
where $\smash{\overline{T}_{\rm cor}}$ is the mean coronal temperature in MK. We show the best fit relation and its central 68\% uncertainty band overlaid on the fitted subsample of data points in purple and orange for {\it XMM-Newton} and eROSITA data in the upper panel of Fig.~\ref{fig:kT-Fx-withfit}. In the lower panel we display the same data points, but data from unresolved multiple systems are represented with open plotting symbols and they are not included in the fit. The fit parameters for all four data sets are given in Table~\ref{tab:kTFxFit}. 

\begin{table}
\centering
\caption{Fit parameters for the $\smash{\overline{T}}_{\rm cor}$- $F_{\!\rm X}$ relation (Eq.~\ref{eq:T_Fx}) shown in Fig.~\ref{fig:kT-Fx-withfit}.}
\label{tab:kTFxFit}
\renewcommand*{\arraystretch}{1.3}
\resizebox{0.49\textwidth}{!}{
\begin{tabular}{l l c c}
\hline\hline
Instrument & \multirow[t]{2}{\widthof{solved multiples?}}{Includes unre- solved multiples?} & $a$ & $b$ \\ \\
\hline
\textit{XMM-Newton} & Y & $0.081^{+0.003}_{-0.003}$ & $0.290^{+0.003}_{-0.003}$\\
eROSITA & Y & $0.190^{+0.024}_{-0.025}$ & $0.242^{+0.011}_{-0.009}$\\
\textit{XMM-Newton} & N & $0.050^{+0.016}_{-0.012}$ & $0.326^{+0.021}_{-0.021}$\\
eROSITA & N & $0.352^{+0.083}_{-0.077}$ & $0.191^{+0.020}_{-0.018}$\\
\hline
\end{tabular}
}
\end{table}

The fits confirm the slight offset identified by eye between the temperature-brightness relation measured with eROSITA and EPIC/pn. The fits for eROSITA are slightly offset and have a shallower slope. 
For a given surface flux eROSITA yields a higher temperature, or for a given temperature eROSITA yields a lower surface flux. Overall, the unresolved multiple stars do not appear to deviate from the single and resolved multiple stars, indicating that our approach to split the observed X-ray flux among the surface area of the components is appropriate. In the fit that excludes the unresolved multiples we find that the difference between the eROSITA and EPIC/pn data persists but the relations are now overlapping within their $1\sigma$ uncertainties. This is mainly due to the loss of the well-constrained data points at the low end of the \textit{XMM-Newton} distribution when $\alpha$\,Cen\,A\&B is removed from the fit. Our analysis indicates that the differences in the instrumental response of the two instruments has a measurable but small impact on the relation.

\subsection{Comparison between high and low resolution spectra}\label{subsubsect:HighLowRes}

\citetalias{Johnstone_2015} discussed the X-ray temperature-brightness relation based on high-resolution X-ray spectra. We show their best fit relation in Eq.~\ref{eq:T_Fx} with values $a\!=\!0.11$ and $b\!=\!0.26$ in black in Figs.~\ref{fig:kT-Fx-Main} and \ref{fig:kT-Fx-withfit}.
Our fit for the EPIC/pn observations of the FGK 10\,pc sample agrees with the relation found by JG15. This agreement is remarkable considering the different ways in which low- and high-resolution X-ray spectra are analysed. In particular, the temperature obtained from high-resolution X-ray spectra derives from a (near)-continuous emission measure distribution, while in low-resolution spectra a model with few discrete temperatures (in our case one to three) is used. Also, more reliable constraints on elemental abundances are extracted from high-resolution spectra while in low-resolution spectra abundances are notoriously degenerate with the emission measure. The excellent match between the X-ray temperature-brightness relation of XMM-Newton's EPIC/pn and RGS implies that the calibration we provide in Table~\ref{tab:kTFxFit} for the FGK 10\,pc sample is not hampered by the limited spectral resolution. The values from Table~\ref{tab:kTFxFit}, or alternatively those by \citetalias{Johnstone_2015}, can therefore be used to estimate the plasma temperature for a given low-luminosity object for which no spectral analysis, not even at low spectral resolution, is possible but the X-ray flux can be determined.

The eROSITA observations place the FGK 10\,pc sample systematically above the relation defined by the \textit{XMM-Newton} data, i.e. the y-axis offset of the fit is higher. An offset to the \citetalias{Johnstone_2015} relation was already noticed by \cite{Magaudda_2022} for a small sample of M dwarfs observed in the eROSITA CalPV field eFEDS. The good agreement between \textit{XMM-Newton} high- and low-resolution data suggests differences in the spectral responses as a cause for the offset with respect to eROSITA. The instrument-specific calibration should, therefore, be used when coronal temperatures are determined from the flux for stars too faint for spectral analysis.

\subsection{Comparison with solar coronal structures}
\label{subsubsect:solarregions}

To implement the aforementioned comparison of the 10\,pc $\smash{\overline{kT}}$-$F_{\!\rm X}$ relation with our Sun, we display in Fig.~\ref{fig:kT-Fx-Main} the range of observed solar values given by \citetalias{Johnstone_2015} (grey dashed line). These values were originally provided by \cite{Peres_2000} in the framework of the {\it Sun as an X-ray star} (SaXS) project in which the solar corona is ``reduced'' to its appearance in a hypothetical observation without spatial resolution, by estimating the surface-averaged X-ray properties of the Sun. Fig.~\ref{fig:kT-Fx-Main} shows that the average Sun covers the lowest part of the flux and temperature range represented by the FGK 10\,pc sample at a slightly smaller slope.

We extend the comparison with the Sun by highlighting the areas covered by the different types of magnetic structures known to be present on the Sun. Most of the region types we consider have been defined within the SaXS project \citep{Orlando_2001, Orlando_2004}. Based on the intensity seen in {\it Yohkoh} \citep{Acton_1992} images, the solar corona was divided into background corona (BKC), active regions (AR), and cores of active regions (CO). An additional typical type of coronal structure on the Sun are coronal holes (CH), regions that are devoid of X-ray emission and often related to open magnetic field lines where plasma can escape into space.
We adopt the ROSAT energy band $F_{\!\rm X}$ values for the solar coronal structures derived by \citet{Caramazza_2023}. To determine the range of plasma temperatures found in each coronal structure, we use their emission measure distributions (EMD) presented by \citet{Joseph_2026} as the range which is covered within one order of magnitude of the peak of their {\it Yohkoh} EMD. If this limit lies between two bins of the EMD, we linearly interpolate the temperature. 
The EMD for CH are not provided by \citet{Joseph_2026}, and therefore we instead use the temperature range from \citet{Heinemann_2021} derived from EUV observations of $707$ solar CH. Since the methodology for deriving the temperatures of CH is different they are not fully consistent with the temperatures for the other coronal structures. However, the accuracy is sufficient for a qualitative comparison.
The ranges of surface X-ray flux and temperature for all types of solar coronal structure that we display in Fig.~\ref{fig:kT-Fx-Main} are listed in Table~\ref{tab:Structures}.

\begin{table}
\centering
\caption{Ranges of temperatures and X-ray surface fluxes of typical solar coronal magnetic structures with values adopted from \citet{Caramazza_2023}, \citet{Joseph_2026}, \citet{Heinemann_2021}.}
\label{tab:Structures}
\begin{tabular}{l c c c c}
\hline\hline
Structure & log $F_{\!\rm X,min}$ & log $F_{\!\rm X,min}$ & $T_{\!\mathrm{min}}$ & $T_{\!\mathrm{max}}$\\
 & $\rm{(erg\,cm^{-2}\,s^{-1})}$ & $\rm{(erg\,cm^{-2}\,s^{-1})}$ & (MK) & (MK)\\
\hline
CH & 3.00 & 3.78 & 0.76 & 1.12\\
BKC & 3.37 & 4.30 & 0.57 & 1.93\\
AR & 4.97 & 5.36 & 0.79 & 4.56\\
CO & 5.89 & 6.86 & 1.44 & 8.26\\
\hline
\end{tabular}
\end{table}

From Fig.~\ref{fig:kT-Fx-Main} we can see that the FGK\,10pc sample stars cover the full range of temperature-brightness combinations defined by the solar coronal structures. Evidently, on a given star at a given time there can be different fractions of the coronal surface covered by different region types, just like observed on the Sun, where e.g. the surface-average ($kT$,$F_{\!\rm X}$) combination during the solar cycle maximum is between that of the regions covered by solar BKC and AR. The comparison of our surface-averaged values for the FGK 10\,pc sample to the solar magnetic regions should, therefore, be taken as a rough measure for the activity level of these stars in relation to our Sun. In this sense, e.g. the absence of any star in the CH region indicates that no star in the FGK\,10pc sample is fully covered by coronal holes. In an \textit{XMM-Newton} observation, \citet{Caramazza_2023} found an upper limit of the X-ray surface flux of the M dwarf GJ\,745\,A (${d\!=\!8.8\,\mathrm{pc}}$) within the range covered by CH. The lack of stars in this range in the FGK\,10pc sample does not exclude that some stars may have a fraction of their surface in a CH-like state, which however contributes only marginally to the X-ray emission. Similarly, on the high-activity end the location of stellar flares in the area of fainter solar region types is due to the fact that flares cover a very small fraction of the corona despite dominating the X-ray output \citep{Joseph_2026}.

\subsection{Identification of possible Maunder minimum stars}
\label{subsubsect:MMstars}

We include in Fig.~\ref{fig:kT-Fx-Main} also the only confirmed star which is known to be in a Maunder minimum-like extended low-activity state, HD\,166620, with values derived by \citet{Bennedik_2026} that place the star in the solar BKC region. At a distance of $11.10\,\mathrm{pc}$ \citep{Gaia_2023}, HD\,166620 is just slightly outside the 10\,pc boundary. Within the FGK\,10pc sample, there are three objects within the parameter range of solar BKC at $\mathrm{log}\,F_{\!\rm X} (\mathrm{erg\,s^{-1}\,cm^{-2}})<4$, namely $\tau$\,Ceti, $\gamma$\,Pavionis, and HD\,20794. Their low X-ray temperature and flux shared with HD\,166620 could make these objects possible Maunder minimum candidates. In Table~\ref{tab:MMdiscussion}, we summarise their SpT, metallicity, rotation periods ($P_\mathrm{rot}$), age estimates, and chromospheric Ca\,II~H\&K flux characterised by the $S_{\rm HK}$-index.

Compared to HD\,166620, the other stars have earlier SpT and are somewhat younger. As far as known, these stars are slow rotators. In fact, $\tau$\,Ceti has a flat $S_{\!{\rm HK}}$ time series \citep{Isaacson_2024} and was proposed as a Maunder minimum candidate \citep[see e.g.][]{Judge_2004}. Its low metallicity, high rotation period, and high age point to an intrinsically low coronal activity instead. The latter also applies for $\gamma$\,Pav, however it has a higher $S_{\!{\rm HK}}$. Without long-term activity monitoring, it is ambiguous whether these stars are intrinsically inactive due to their high age or are truly in an extended cycle minimum. Only for HD\,20794, \citet{Nari_2025} report evidence for an $\sim \!\!8$\,yr cycle from various activity indicators. Its S-index amplitude is, however, very low.

\subsection{Impact of X-ray variability on coronal temperatures}\label{subsect:XrayVar}

\begin{figure*}
    \centering
    \includegraphics[width=0.8\textwidth,trim={1.15cm 1.47cm 1.82cm 2.25cm},clip]{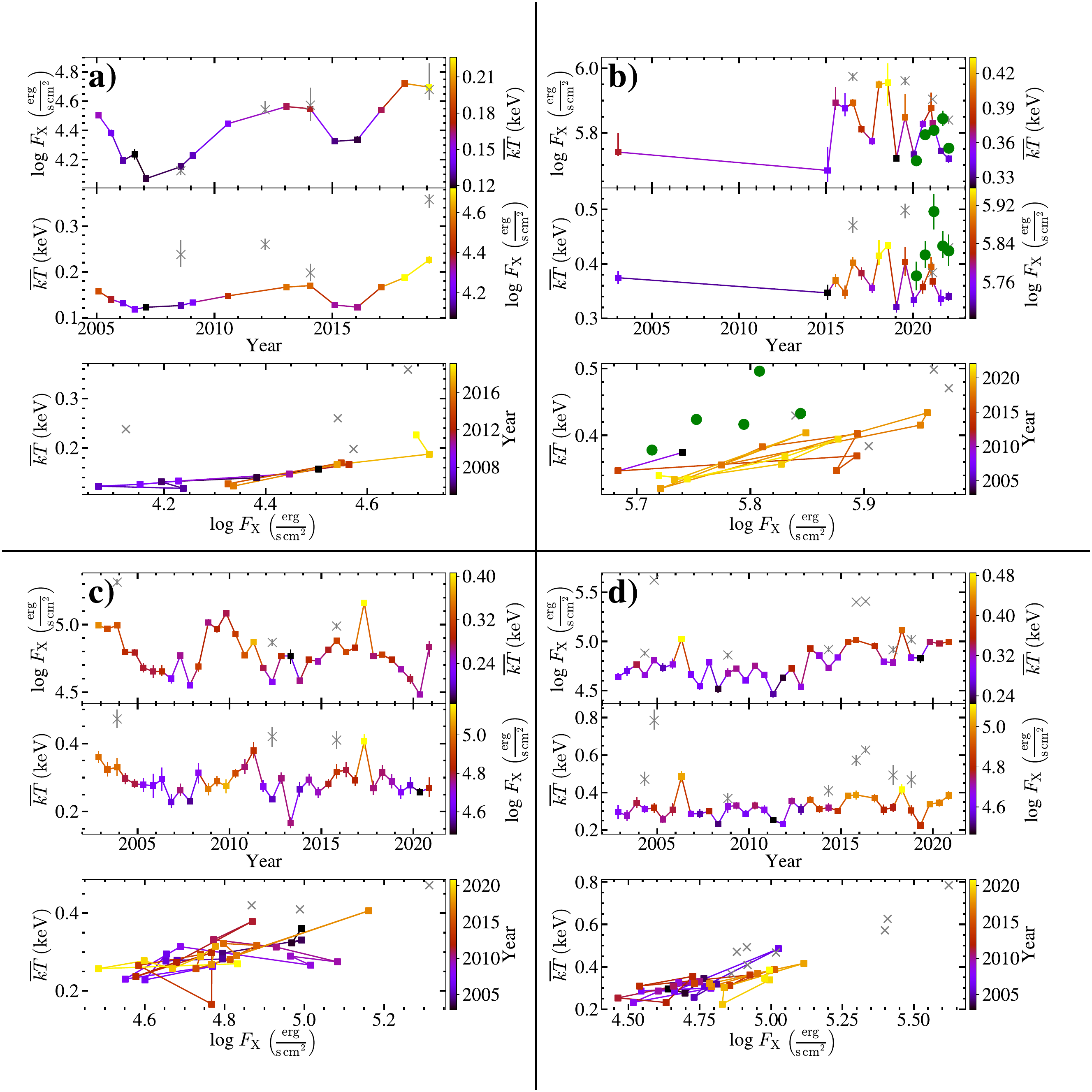}
    \caption{Time series of spectral parameters from \textit{XMM-Newton} observations (squares) for \textbf{a)} $\alpha$\,Cen\,A\&B, \textbf{b)} $\epsilon$\,Eri, \textbf{c)} 61\,Cyg\,A, and \textbf{d)} 61\,Cyg\,B. Flares are shown with grey crosses. eRASS detections of $\epsilon$\,Eri are shown with green circles. \textit{Upper panels:} X-ray surface flux. \textit{Middle panels:} Mean coronal temperature. \textit{Lower panels:} Evolution of $\smash{\overline{kT}}$-$F_{\!\rm X}$. Error bars not shown in lower panels for clarity.
    }
    \label{fig:Timelines_All}
\end{figure*}

A few stars in the FGK\,10pc sample have numerous observations because they had dedicated {\it XMM-Newton} observing programs with the aim to search for an X-ray activity cycle. Specifically, this regards $\alpha$\,Cen\,A\&B \citep[see e.g.][]{Ayres_2023, Robrade_2012}, 61\,Cyg\,A\&B \citep{Robrade_2012} and $\epsilon$\,Eri \citep{Coffaro_2020, Fuhrmeister_2023}. Their X-ray cycles were identified based on the variation of their X-ray luminosities. Here, we add a study of the evolution of coronal temperature and its relation to the evolution of coronal brightness.
In Fig.~\ref{fig:Timelines_All} we display for each of these systems the time series of $F_{\!\rm X}$, $\smash{\overline{kT}}$, and their relation, the latter being a zoomed version of the temperature-brightness relation from Fig.~\ref{fig:kT-Fx-Main}. Data points are connected according to their chronological order. Flares identified by us and analysed separately from the quiescent emission are highlighted as grey data points. In the following we discuss the results for the individual objects.

The binary $\alpha$\,Cen\,A\&B is unresolved with {\it XMM-Newton}'s EPIC/pn and therefore, we treat it as a single source. Special data treatment involving a customised photon extraction using the smaller PSFs of the MOS detectors by \cite{Robrade_2012} showed that the X-ray light is dominated by $\alpha$\,Cen\,B. Furthermore, $\alpha$\,Cen\,A has a low cycle amplitude \citep{Ayres_2023}. While the cyclic behaviour of $F_{\!\rm X}$ was expected considering the previous studies of these data, we demonstrate in the middle panel of Fig.~\ref{fig:Timelines_All}~a) that the X-ray temperature follows the same behaviour, that is the corona is getting hotter during the cycle maxima and cooler during the minima. The flares have low amplitudes when compared to the other stars discussed in this section and thus they cause a strong increase of temperature but not $F_{\!\rm X}$.

The K2 dwarf $\epsilon$\,Eri was observed $17$ times with \textit{XMM-Newton}. It is located in the western Galactic sky and therefore, it provides an excellent cross-calibration point for eROSITA and EPIC/pn. Although the observations with both instruments are not strictly simultaneous, the multi-epoch data available for both data sets allows us to examine systematic trends. The eROSITA data (green circles in Fig.~\ref{fig:Timelines_All}~b)) clearly follows the cyclic trend where the time-evolution of flux mimics that seen with EPIC/pn. The comparison of eRASS data with the EPIC/pn time series in the top and middle panels of Fig.~\ref{fig:Timelines_All}~b) illustrates the enhanced temperature measured by eRASS with respect to EPIC/pn at similar, or even lower, flux level. This manifests in the offset of the two fit relations discussed in Sect.~\ref{subsect:Powerlaw}. We defer a more detailed study of the combined {\it XMM-Newton} and eROSITA data base for the X-ray cycle of $\epsilon$\,Eri to a future work.

The binary 61\,Cyg\,A\&B is resolved with EPIC/pn. Therefore, we present the results for 61\,Cyg\,A\&B individually in Fig.~\ref{fig:Timelines_All}~c) and d). While 61\,Cyg\,B does not show an unambiguous activity cycle \citep{Ayres_2025}, the cycle of 61\,Cyg\,A presented by \citet{Robrade_2012} and \citet{Ayres_2025} is reproduced by the coronal temperature. Both stars have exhibited high-amplitude flares. This explains the much higher X-ray surface fluxes seen during their flares when compared with the flares of $\alpha$\,Cen\,A\&B.

\section{Physical interpretation of the temperature-brightness relation}\label{sect:RTV}

\subsection{Explanation with the Rosner-Tucker-Vaiana scaling law}
\label{sec:rtv_derivation}
In this section, we show that the slope of the empirical power law representing the X-ray temperature-brightness relation discussed in Sect.~\ref{sect:EmpiricalRelation} can be derived from fundamental assumptions of the underlying physics of the stellar coronae expressed in one of the Rosner-Tucker-Vaiana (RTV) scaling laws \citep{Rosner_1978}, namely the relation between coronal loop length, loop temperature, and loop pressure. These scaling laws result from the assumption that the thermodynamic properties of a coronal loop are governed by the balance between volumetric heating, thermal conduction, and radiative losses. In the following we provide an order-of-magnitude argument which illustrates that the RTV scaling laws naturally predict the empirically observed slope of the coronal temperature-brightness relation. 

For fully ionised coronal plasma, the X-ray luminosity scales with the emission measure, $EM$, as
\begin{equation}
    L_{\mathrm{X}}\approx EM\cdot\Lambda(T_{\mathrm{cor}}),
\end{equation}
where $\Lambda(T_{\mathrm{cor}})$ is the temperature dependent coronal cooling function. If we assume that coronal loops with a characteristic length $L$ occupy a surface filling factor $f$ of the stellar corona, the definition of the emission measure can be expressed by
\begin{equation}
    EM = \int n_{\mathrm{e}}^2\,\mathrm{d}V \approx n_{\mathrm{e}}^2 \cdot 4\pi R^2 \cdot f \cdot L,
\end{equation}
with $n_{\mathrm{e}}$ as the electron particle density. Using the ideal gas equation ${n_{\mathrm{e}}\propto p/T}$ where $p$ is the pressure, we can rearrange to obtain the X-ray surface flux
\begin{equation}\label{eq:FxRTV}
    F_{\!\rm X}\propto \frac{EM}{4\pi R^{2}} \cdot \Lambda(T_{\mathrm{cor}}) \approx f\cdot \left(\frac{p}{T_{\mathrm{cor}}}\right)^{2} \cdot L \cdot \Lambda(T_{\mathrm{cor}}).
\end{equation}
We approximate the cooling function as a power law, ${\Lambda(T_{\mathrm{cor}})\propto T_{\mathrm{cor}}^{-m}}$, where the value of $m$ depends on the coronal elemental abundances and temperature. We justify this approximation in Sect.~\ref{subsubsect:LossFunction}. The RTV scaling law states that ${T_{\rm cor}\propto(p\cdot L)^{1/3}}$. Inserting both relations in Eq.~\ref{eq:FxRTV} gives an expression that relates $T_{\mathrm{cor}}$ and $F_{\!\rm X}$:
\begin{equation}\label{eq:t_cor}
    T_{\mathrm{cor}} \propto \left(\frac{F_{\!\rm X}\cdot L}{f}\right)^\frac{1}{4-m}.
\end{equation}
For a fixed loop length and coronal filling factor the exponent represents the fit parameter $b$ in Eq.~\ref{eq:T_Fx}. For a wide range of coronal temperatures between 0.4--30\,MK, $m$ has values approximately between 0.2--0.8 \citep[see e.g.][]{Raymond_1976, Raymond_1977, Orlando_2005}. We discuss the cooling function and its power law approximations in Sect.~\ref{subsect:CoolingFunction}. For ${m=\left[0.2,0.5,0.8\right]}$ this yields an exponent of ${b=\left[0.26,0.29,0.31\right]}$ respectively. These values are in remarkable agreement with the empirical slopes we obtained for the FGK 10pc sample (see Table~\ref{tab:kTFxFit}) and that were found in previous studies of the X-ray temperature-brightness relation \citepalias{Johnstone_2015, Binder_2026}. In this framework, the scatter in the observed distribution can be explained by varying loop lengths, filling factors, and coronal abundances between the different stars. 

We emphasise that the predicted temperature-brightness relation depends only weakly on the value of $m$. Even when varying $m$ between the two extreme cases considered here, the exponent of the resulting power law changes by less than 20\%. Thus, although we adopt a crude approximation for the cooling function, the resulting uncertainty in the predicted slope is small. The close agreement between the predicted and observed slopes suggests that the empirical temperature-brightness relation is largely governed by the RTV scaling law linking pressure, temperature and length of the coronal loop, while the detailed shape of the radiative loss function introduces only a secondary correction. In this interpretation, the observed relation represents the statistical manifestation of the underlying physics of magnetically confined structures in the stellar coronae.

\subsection{Explanation for the outliers}\label{subsect:RTV_Outliers}

By visual inspection, we have identified in Fig.~\ref{fig:kT-Fx-Main} the four outliers, 61\,Cyg\,A\&B, GJ\,53\,A, and GJ\,783\,A. Their observed offset from the well-defined temperature-brightness relation formed by the majority of the stars cannot be explained by instrumental differences since we are only comparing \textit{XMM-Newton} observations. Since 61\,Cyg\,A\&B have multi-epoch data, we can measure the slope of the temperature-brightness relation separately for these stars alone and find an interpretation in terms of the RTV scaling law.

\begin{figure}
    \centering
    \includegraphics[width=0.45\textwidth]{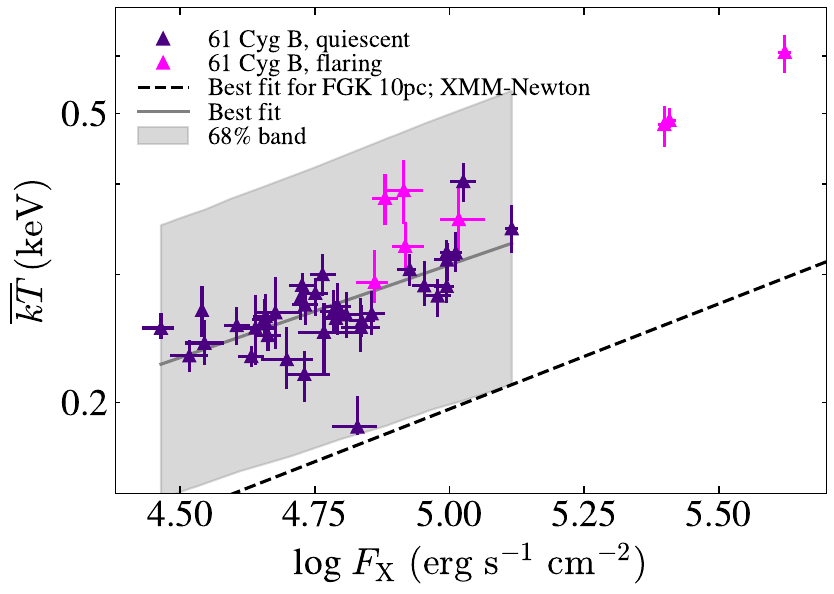} 
    \caption{X-ray temperature-brightness relation for \textit{XMM-Newton} observations of 61\,Cyg\,B, best fit over its quiescent periods (grey), and fit for EPIC/pn observations of FGK\,10pc sample from Fig.~\ref{fig:kT-Fx-withfit} (black, dashed).}
    \label{fig:61CygFit}
\end{figure}

Fitting for the outlier star 61\,Cyg\,B, analogously to our approach for the main sample in Sect.~\ref{subsect:Powerlaw}, only the \textit{XMM-Newton} observations representing the quiescent state (see Fig.~\ref{fig:61CygFit}), we find an offset of ${a=0.19^{+0.07}_{-0.05}}$ and a slope of ${b=0.26^{+0.03}_{-0.03}}$. While the slope is consistent with the fit of the EPIC/pn observations of the bulk of the FGK\,10pc sample listed in Table~\ref{tab:kTFxFit}, the offset is substantially higher. To explain this offset, we can express Eq.~\ref{eq:t_cor} in its double-logarithmic form:
\begin{equation}
     \log T_{\mathrm{cor}} \propto \frac{1}{4-m} \cdot \log F_{\!\rm X} + \frac{1}{4-m} \cdot \log\left( \frac{L}{f}\right),
    \label{eq:logt_cor}
\end{equation}
where it becomes obvious that the ratio $L/f$ defines the offset for a given slope. At a fixed X-ray surface flux, we obtain
\begin{equation}\label{eq:offsets}
    \frac{(L/f)_{\rm 61\,Cyg\,B}}{(L/f)_{\rm bulk\,FGK}} = \left(\frac{T_{\rm 61\,Cyg\,B}}{T_{\rm bulk\,FGK}}\right)^{1/b},
\end{equation}
ignoring the slight differences in the observed slope $b$, and hence $m$. At the observed values of $F_{\!\rm X}$, the coronal temperatures of 61\,Cyg\,B imply an approximately 4--7 times larger ratio $L/f$ for the star than for the bulk of the FGK\,10pc sample, considering the range of value for $m$ from Sect.~\ref{sec:rtv_derivation} and Fig.~\ref{fig:CoolingFunction}. This indicates that 61\,Cyg\,B either has longer coronal loops or a smaller filling factor. 61\,Cyg\,B frequently exhibits high-amplitude flares, of which an example is shown in Fig.~\ref{fig:LC_Flare}. As more active stars exhibit larger coronal filling factors \citep[see e.g.][]{Coffaro_2020, Drake_2023}, it is unlikely that the offset of 61\,Cyg\,B could be explained with a smaller value of $f$. Thus, evidence points towards increased loop lengths when compared to the other stars within the FGK\,10pc sample. 

For 61\,Cyg\,A, the best-fit $\smash{\overline{kT}}$-$F_{\!\rm X}$ pairs yield a significantly shallower slope ${b=0.17^{+0.03}_{-0.03}}$ and a high offset ${a=0.44^{+0.16}_{-0.12}}$. The slope would require an unphysical negative value for $m$. A possible interpretation could be a change of the offset over time due to a changing ratio $L/f$. With Eq.~\ref{eq:offsets} and an assumed ${b=0.29}$, we compute an approximately 3--5 times larger ratio $L/f$ for 61\,Cyg\,A compared to the FGK\,10pc sample at the upper end at ${\log F_{\!\rm X}\,\mathrm{(erg\,cm^{-2}\,s^{-1})} = 5.10}$ and at the lower end of the relation with ${\log F_{\!\rm X}\,\mathrm{(erg\,cm^{-2}\,s^{-1})} = 4.50}$, respectively.

GJ\,783\,A and GJ\,53\,A each have only one \textit{XMM-Newton} observation (ObsIDs 0670380401 and 0982040301) and are in the Russian eROSITA sky half. The EPIC/pn light curve of GJ\,783\,A contains one short low-amplitude flare which was removed as described in Sect.~\ref{subsubsect:LCs}. The X-ray spectra of both systems have a marked high-energy tail and therefore their high coronal temperatures are physical and not a result of the fitting procedure. Analogous to 61\,Cyg\,B, a different value for $L/f$ could offer a possible explanation for their offset from the global temperature-brightness relation. Alternatively, it is possible that the X-ray emission of GJ\,783\,A and GJ\,53\,A is dominated by their unresolved M dwarf companions. In fact, computing $F_{\!\rm X}$ with exclusively the M dwarf surface areas would put both systems in agreement with the $\smash{\overline{kT}}$-$F_{\!\rm X}$ relation shown in Fig.~\ref{fig:kT-Fx-Main}. GJ\,53\,A and GJ\,783\,A are old with estimated ages of $12.7\pm2.7\,\mathrm{Gyr}$ \citep{Bond_2020} and $7.7\,\mathrm{Gyr}$ \citep{Mamajek_2008} respectively, which makes it plausible that the earlier-type stars in the binary are X-ray faint. As part of a \textit{Chandra} monitoring program (PI: Binder), GJ\,783 was recently observed. Due to the excellent spatial resolution of \textit{Chandra}, analysis of these observations will allow testing this hypothesis in the near future.

\subsection{Assumptions and limitations of the RTV-based derivation}\label{subsect:CoolingFunction}
The derivation presented in Sect.~\ref{sec:rtv_derivation} is an order-of-magnitude argument to illustrate that the observed temperature-brightness relation naturally follows from the scaling laws governing magnetically confined plasma structures of the corona. It is not intended as a rigorous derivation for a realistic corona, which consists of a multi-thermal ensemble of magnetic structures spanning a broad range of sizes, temperatures, heating conditions, etc. In the following, we discuss the main assumptions and limitations underlying the derivation and its application to the data. 

\subsubsection{Approximation of the radiative loss function}\label{subsubsect:LossFunction}
Our parametrisation of the radiative loss function of the corona as a power law in temperature is only a coarse description of the actual cooling function which changes continuously with temperature \citep[e.g.][]{Raymond_1976, Raymond_1977}. As a result, there is no unique value of the exponent $m$ in the approximation $\Lambda(T) \propto T^{-m}$.

\begin{figure}
    \centering
    \includegraphics[width=0.45\textwidth,trim={2.7cm 13.2cm 1.85cm 5.1cm},clip]{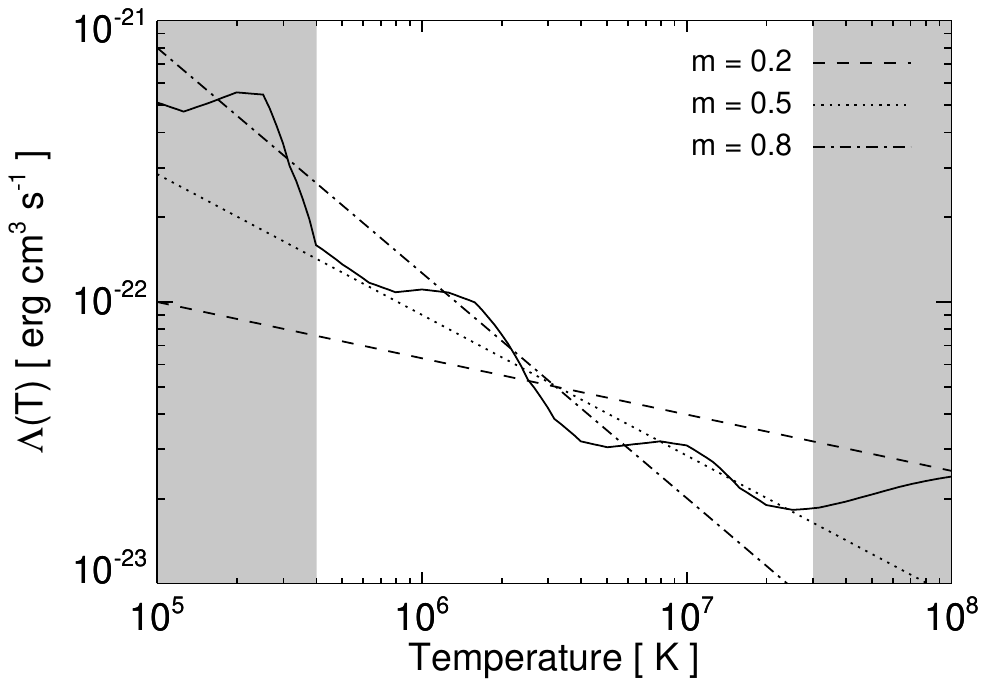}
    \caption{Coronal cooling function (solid line) and approximations with power laws with varied exponents.
    }
    \label{fig:CoolingFunction}
\end{figure}

Fig.~\ref{fig:CoolingFunction} compares the radiative loss function with three representative power law approximations. The temperature range ${[4\cdot 10^5, 3\cdot 10^7]\,\mathrm{K}}$ relevant for our study, that is the range which includes the coronal temperatures of the FGK stars in our sample, is represented by the white area. By visual inspection the case ${m = 0.5}$ provides a reasonable description of the cooling function across the entire temperature range. The values ${m=0.2}$ and ${m=0.8}$ were considered as representative lower and upper limits that approximately bracket the cooling function over the same range. These values, that we associated in Sect.~\ref{sec:rtv_derivation} with the observed slope in the temperature-brightness relation, are therefore not the result of a formal fit, but merely illustrate the range of plausible effective slopes that may be adopted for the analytical approximation of the cooling function. 

\subsubsection{Applicability of the RTV scaling law}\label{subsubsect:RTV_Applicability}

The RTV scaling law relates the maximum temperature of an individual static coronal loop to its pressure and length (see Sect~\ref{sec:rtv_derivation}). By contrast, the temperature in the observed temperature-brightness relation is measured from unresolved stellar X-ray spectra and, therefore, it is an emission measure weighted average over the entire stellar corona. Clearly this does not correspond to the apex temperature of any individual loop composing the corona.

The comparison between the empirical temperature-brightness relation and the RTV prediction should, therefore, be considered as statistical. In fact, it implies that the average properties of the dominant X-ray emitting loop population can be represented by characteristic values of loop pressure and length.

\subsubsection{Loop length and filling factor}
Under the simplifying assumption that the characteristic loop length and filling factor do not vary systematically, Eq.~\ref{eq:t_cor} derived in Sect.~\ref{sec:rtv_derivation} reduces to a universal relation
\begin{equation}
{T_{\mathrm{cor}} \propto ({F_{\!\rm X}})^{1/(4-m)}},
\end{equation}
while large independent variations of the characteristic loop length and the filling factor would be expected to produce a substantial spread in Fig.~\ref{fig:kT-Fx-Main}.

The coronal filling factor, defined here as the fraction of the surface covered with magnetic loops, can also be identified with the fractional coronal surface covered with the different types of magnetic structures inferred from analogy with the Sun. These can be measured in the framework of the SaXS methodology (see Sect.~\ref{subsubsect:solarregions}). In such studies it was found that more active stars exhibit larger coronal filling factors \citep{Coffaro_2022, Drake_2023}. At the same time, flare analyses indicate that late-type stars can host magnetic loops spanning a broad range of sizes, from compact structures to loops comparable to or even exceeding the stellar radius \citep[e.g.][]{Favata_2005, Mitra_Kraev_2005, Reale_2014}.
However, similar measurements are generally not feasible for the much fainter quiescent coronal structures that dominate the X-ray emission of most stars. The characteristic loop length entering our analytical derivation should, therefore, be regarded as an effective parameter describing the dominant X-ray-emitting loop population rather than the size of individual magnetic structures. Consequently, deviations from the power law that best fits the observed temperature-brightness relation are expected and they likely reflect genuine differences in coronal magnetic structure rather than observational uncertainties alone.

The relatively small intrinsic scatter, therefore, suggests that the spread of the quantity $(L/f)$ entering the analytical derivation remains confined to a relatively narrow range across the FGK\,10pc sample with the exception of the ``outliers'' marked in grey in Fig.~\ref{fig:kT-Fx-Main} and discussed in Sect.~\ref{subsect:RTV_Outliers}. This does not necessarily imply that individual loop lengths or filling factors are nearly the same on all stars and constant in time. Rather, it indicates that the characteristic properties of the X-ray-emitting loop population are not independent but correlated among each other, such that differences in loop length and filling factor largely compensate each other. Specifically, more active stars (characterized by higher $T_{\rm cor}$ and higher $F_{\rm X}$) that are generally known to have larger coronal filling factors also must have larger loop lengths in the dominating type of coronal structure.

The above interpretation is supported by the results of the SaXS analysis presented by \citet{Coffaro_2022}, where the very active star Kepler-63 required a corona with a significant extension above the photosphere to keep the filling factor within the limit of $100\%$ surface coverage. Within the framework proposed here, this naturally suggests that stars with larger effective filling factors also possess larger characteristic coronal loop lengths, corresponding to more vertically extended coronae. The consistency between these independent observational results supports the idea that the stellar dynamo regulates not only the total amount of magnetic flux emerging at the stellar surface, but also the characteristic geometry of the X-ray-emitting coronal structures. The changes in loop size and surface coverage associated with variations of the activity level appear to occur in a correlated way, maintaining a nearly invariant effective ratio $(L/f)$.

\section{Summary and conclusions}\label{sect:summary}
We have constructed a comprehensive X-ray catalogue for the volume-limited sample of the FGK-type stars within 10\,pc. Unlike previous studies, we do not rely on rate-to-flux conversion factors because we perform spectral analysis on all detections with sufficient count numbers with \textit{XMM-Newton} and eROSITA. Together with careful handling of multiplicity and optical loading of the X-ray detector, this allowed us to build a homogeneous catalogue of X-ray fluxes and coronal temperatures derived from spectral fitting. This data set allows us to systematically address instrumental aspects relevant in the analysis of stellar X-ray data and to sample a broad range of coronal properties.

For the mean coronal temperature $\smash{\overline{kT}}$, calculated from a logarithmic weighting with emission measures, we confirm previous evidence \citepalias{Johnstone_2015, Binder_2026} for a power law relation with the surface X-ray flux. We find different values for the power law parameters, slope and offset, between the eRASS and the EPIC/pn data sets, quantifying previous suspicions by \citet{Magaudda_2022} that this empirical law is instrument-specific. Interestingly, the relation we derive from the low-resolution EPIC/pn data is in excellent agreement with the one obtained by \citetalias{Johnstone_2015} from high-resolution X-ray observations, establishing that detailed modelling of the emission measure distribution is not necessary to retrieve the average coronal properties of stars.

The temperature-brightness relation of the FGK\,10pc sample closely follows the ranges of temperatures and X-ray surface fluxes observed in solar coronal structures, implying that their properties are universal across late-type main sequence stars. The low end of the relation is consistent with the Maunder minimum star HD\,166620 \citep{Bennedik_2026}. Three stars in the FGK\,10pc sample have similar properties. While evidence for a cycle was found for one star, the other two stars could potentially be in a Maunder minimum state. Further long-term monitoring of their chromospheric activity is required to unambiguously determine whether they are in such a Grand Minimum.

We present the first dedicated study of the joint variability of X-ray temperature and brightness. For three systems with long-term \textit{XMM-Newton} observing programs, the coronal temperature closely follows the evolution of the X-ray flux over the course of the X-ray cycle which was identified in previous literature through the variation of their X-ray luminosity. This indicates as the origin for the X-ray cycle a periodic change of the coronal filling factors of individual types of solar-like magnetic structures as suggested by \citet{Joseph_2026}.
Analogous to the Sun, the coronae appear to be dominated by BKC during activity cycle minima and towards the cycle maxima the contribution from AR and CO increases. Furthermore, flares are more prevalent near cycle maxima and lead to transient increases in coronal temperatures as discussed in Sect.~\ref{subsect:XrayVar}. 

We showed that the temperature-surface flux relation can be derived from the RTV scaling law. This suggests that the observed relation reflects the fundamental physics governing magnetically confined coronal plasma. Fig.~\ref{fig:kT-Fx-Main} provides an important observational link supporting this interpretation. The various classes of solar coronal structures (including CH\footnote{CH are dominated by open magnetic fields, which are not described by the RTV scaling laws. However, their X-ray emission arises from embedded small-scale closed loops that are approximately in hydrostatic equilibrium and thus expected to follow RTV-like scaling \citep[e.g.][]{Hudson_2002, Cranmer_2009}. Their lower position in the $kT$-$F_{\rm X}$ diagram reflects the reduced filling factor and pressure of these structures.}, BKC, AR, and CO) occupy the same temperature-surface flux plane as the stellar sample. Since these structures are all composed of magnetic loops that approximately obey the RTV scaling law, their location along the observed relation of global parameters indicates that the same underlying physical processes regulate plasma pressure, temperature, and magnetic geometry over a wide range of spatial scales and activity levels.

The position of an unresolved star in the $kT$-$F_{\rm X}$ diagram is determined by the integrated emission from all magnetic structures present in its corona. Consequently, the observed star-to-star variations do not necessarily imply fundamentally different types of stellar coronae. Instead, they can be interpreted as reflecting different statistical mixtures of solar-like magnetic structures \citep[see also][]{Orlando_2001, Orlando_2004, Orlando_2017}. In relatively inactive stars, the X-ray emission is expected to be dominated by quiet-corona and network-like loops, whereas progressively more active stars become increasingly dominated by AR and CO \citep[e.g.][]{Coffaro_2020, Coffaro_2022}. The resulting increase in plasma pressure, filling factor, and loop size shifts the integrated coronal properties toward higher temperatures and larger X-ray surface fluxes while preserving the same underlying RTV scaling.

As we presented in Fig.~\ref{fig:kT-Fx-Main}, the temperature-brightness relation from more sensitive \textit{XMM-Newton} observations extends towards lower luminosities and temperatures compared to eROSITA. Similarly, future more sensitive instrumentation like the \textit{NewAthena} mission \citep{Cruise_2024} will improve the statistics at the low end of the relation and enable an extension to even fainter coronal emitters, such as brown dwarfs.

\section*{Data availability}
Tables \ref{tab:FundamentalProperties} and \ref{tab:IndividualObservations} are only available in electronic form at the CDS via \url{http://cdsweb.u-strasbg.fr/cgi-bin/qcat?J/A+A/999/A999}.

\begin{acknowledgements} 
During the writing of this paper, we lost our dear colleague Jan Robrade, who made substantial contributions to the eROSITA mission. He is greatly missed by many.
MB acknowledges support by the Bundesministerium für Forschung, Technologie und Raumfahrt through the Deutsches Zentrum für Luft- und Raumfahrt (DLR) under grant number FKZ 50 OR 2505.
This work is based on observations obtained with \textit{XMM-Newton}, an ESA science mission with instruments and contributions directly funded by ESA Member States and NASA; and of archival data of the ROSAT space mission.
This work is also based on data from eROSITA, the soft X-ray instrument aboard SRG, a joint Russian-German science mission supported by the Russian Space Agency (Roskosmos), in the interests of the Russian Academy of Sciences represented by its Space Research Institute (IKI), and DLR. The SRG spacecraft was built by Lavochkin Association (NPOL) and its subcontractors, and is operated by NPOL with support from the MPE. The development and construction of the eROSITA X-ray instrument was led by MPE, with contributions from the Dr. Karl Remeis Observatory Bamberg \& ECAP (FAU Erlangen-Nürnberg), the University of Hamburg Observatory, the Leibniz Institute for Astrophysics Potsdam (AIP), and the Institute for Astronomy and Astrophysics of the University of Tübingen, with the support of DLR and the Max Planck Society. The Argelander Institute for Astronomy of the University of Bonn and the Ludwig Maximilians Universität Munich also participated in the science preparation for eROSITA.
The eROSITA data shown here were processed using the eSASS software system developed by the German eROSITA consortium.
This research has made use of data and software provided by the High Energy Astrophysics Science Archive Research Center (HEASARC), which is a service of the Astrophysics Science Division at NASA/GSFC.
This work has made use of data from the ESA mission {\it Gaia} (\url{https://www.cosmos.esa.int/gaia}), processed by the {\it Gaia} Data Processing and Analysis Consortium (DPAC, \url{https://www.cosmos.esa.int/web/gaia/dpac/consortium}). Funding for the DPAC has been provided by national institutions, in particular the institutions participating in the {\it Gaia} Multilateral Agreement. 
\end{acknowledgements}

   \bibliographystyle{aa}
   \bibliography{FGK_biblio}

\begin{appendix} 

\section{Special cases in source extraction}\label{app:SpecialExtractions}
Two stars required a customised definition of the source extraction due to partial blending with a companion star in a multiple system. In the following, we present both cases in detail.
\subsection{GJ 570 A}
\begin{figure}[h!]
    \centering
    \includegraphics[width=0.45\textwidth]{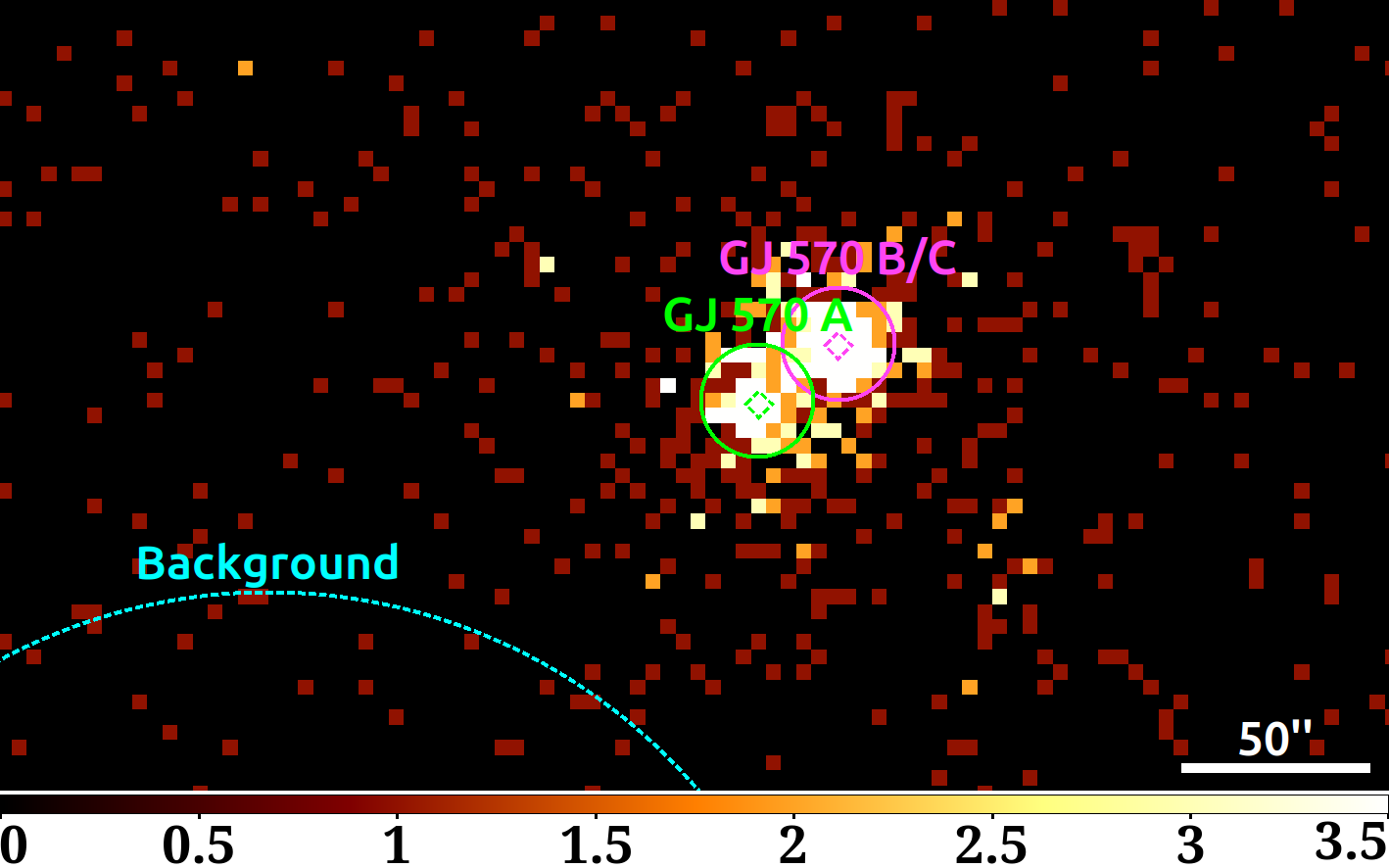}
    \caption{Source (green) and background (cyan, dashed) regions used for extracting the eRASS:5 spectrum and light curve of GJ\,570\,A centred on the positions from source detection. The extent of the PSF of its companions is shown in magenta and diamond symbols show the position of the optical counterparts after correcting for proper motion to the mean eRASS:5 epoch. The colour bar codes the number of counts in each pixel.
    }
    \label{fig:GJ570}
\end{figure}

In the individual eRASS, the GJ\,570 system was observed with too few counts to separate GJ\,570\,A from its two M dwarf companions GJ\,570\,B\,\&\,C. However, two separate sources are detected eRASS:5 as depicted in Fig.~\ref{fig:GJ570}. While the PSFs overlap, the separation is large enough for extraction of individual spectra and light curves. We manually select the source and background regions.

\subsection{GJ 166 A}
\begin{figure}[h!]
    \centering
    \includegraphics[width=0.45\textwidth]{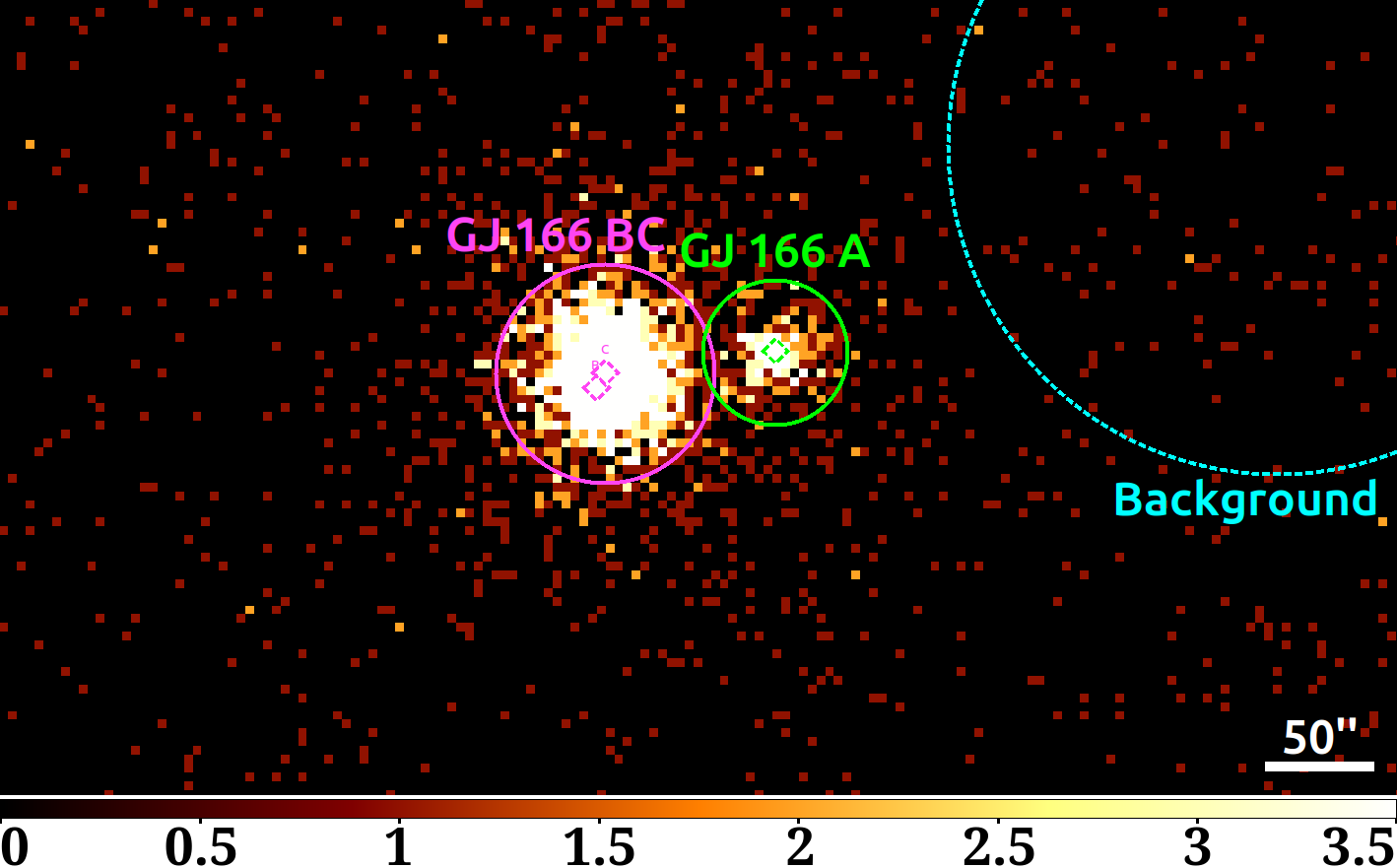}
    \caption{Same as Fig.~\ref{fig:GJ570}, but for GJ\,166\,A.
    }
    \label{fig:GJ166}
\end{figure}

While GJ\,166\,A is is detected as a separate source from its companions, the extraction region automatically defined by the pipeline include both sources in each survey. The companions are a binary with a white dwarf, which is not expected to emit X-ray radiation, and an M dwarf. Thus, we manually define the extraction regions as presented in Fig.~\ref{fig:GJ166}.

\section{Sources affected by optical loading}\label{app:OpticalLoading}
\subsection{Spectra with contributions from optical loading}\label{app:LoadingSpectra}

\begin{figure}[h!]
    \centering
    \includegraphics[width=0.48\textwidth]{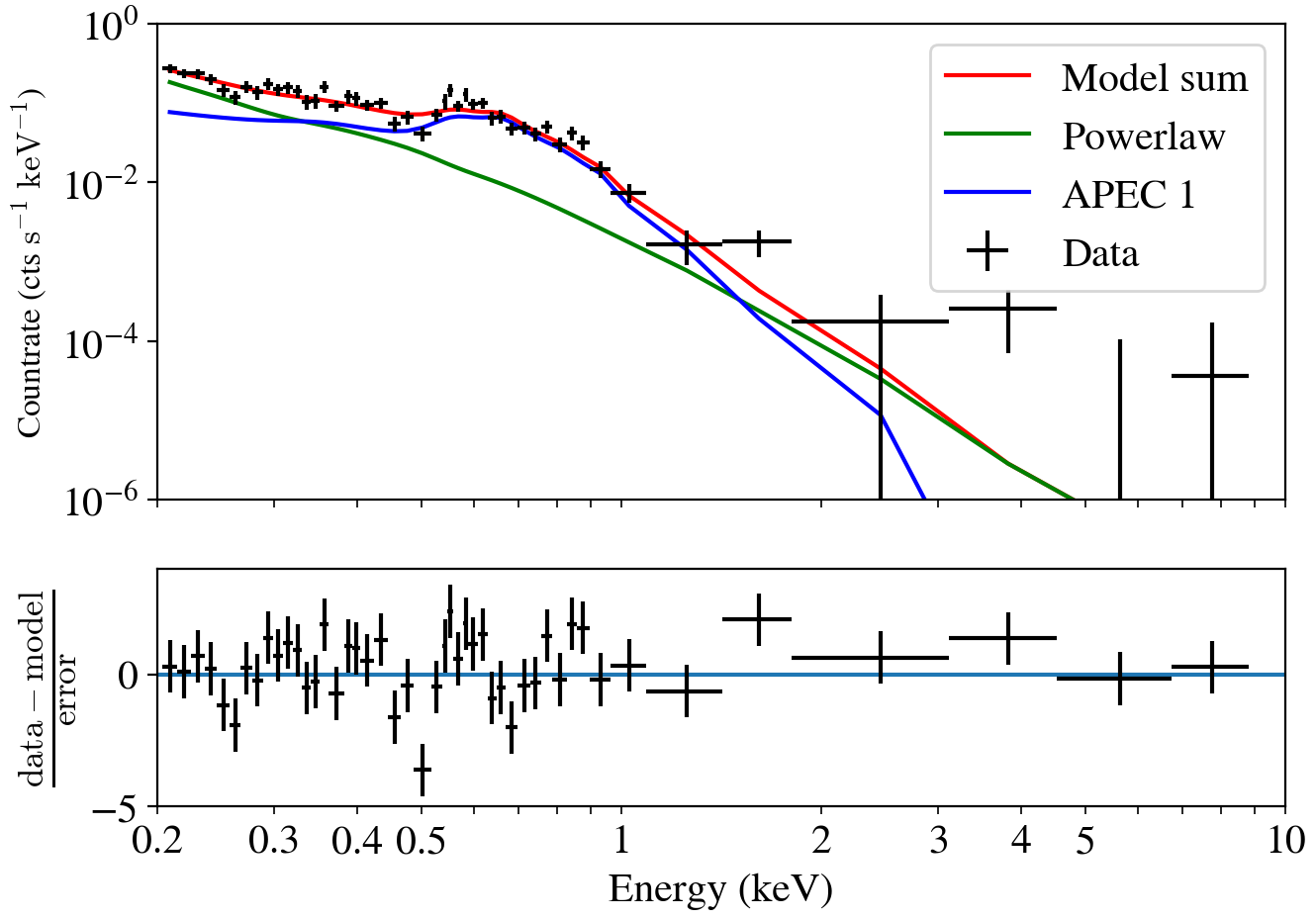}
    \caption{\textit{Upper panel:} \textit{XMM-Newton} EPIC/pn spectrum of HD\,219134 and model fit with power law and 1T-APEC components. \textit{Lower panel:} residuals.
    }
    \label{fig:SpecHD219134}
\end{figure}

\paragraph{HD 219134}
As the optically brightest star in the FGK\,10pc sample that was observed with \textit{XMM-Newton} (ObsID: 0784920201) using the \textsc{medium} filter, the K3-dwarf HD\,219134 is affected by optical loading. In Fig.~\ref{fig:SpecHD219134}, we present the X-ray spectrum together with a model fit containing a power law and a single-temperature APEC-component. While a two-temperature APEC model without a power law component yields a slightly lower $\chi^2_{\mathrm{reduced}}$, the cooler APEC component has a very low temperature and is badly constrained. Thus, we decide that the model containing a power law is a more appropriate description of the spectrum; with the caveat that the intrinsic coronal temperature and X-ray flux of the star are not well constrained.

\begin{figure}[h!]
    \centering
    \includegraphics[width=0.48\textwidth]{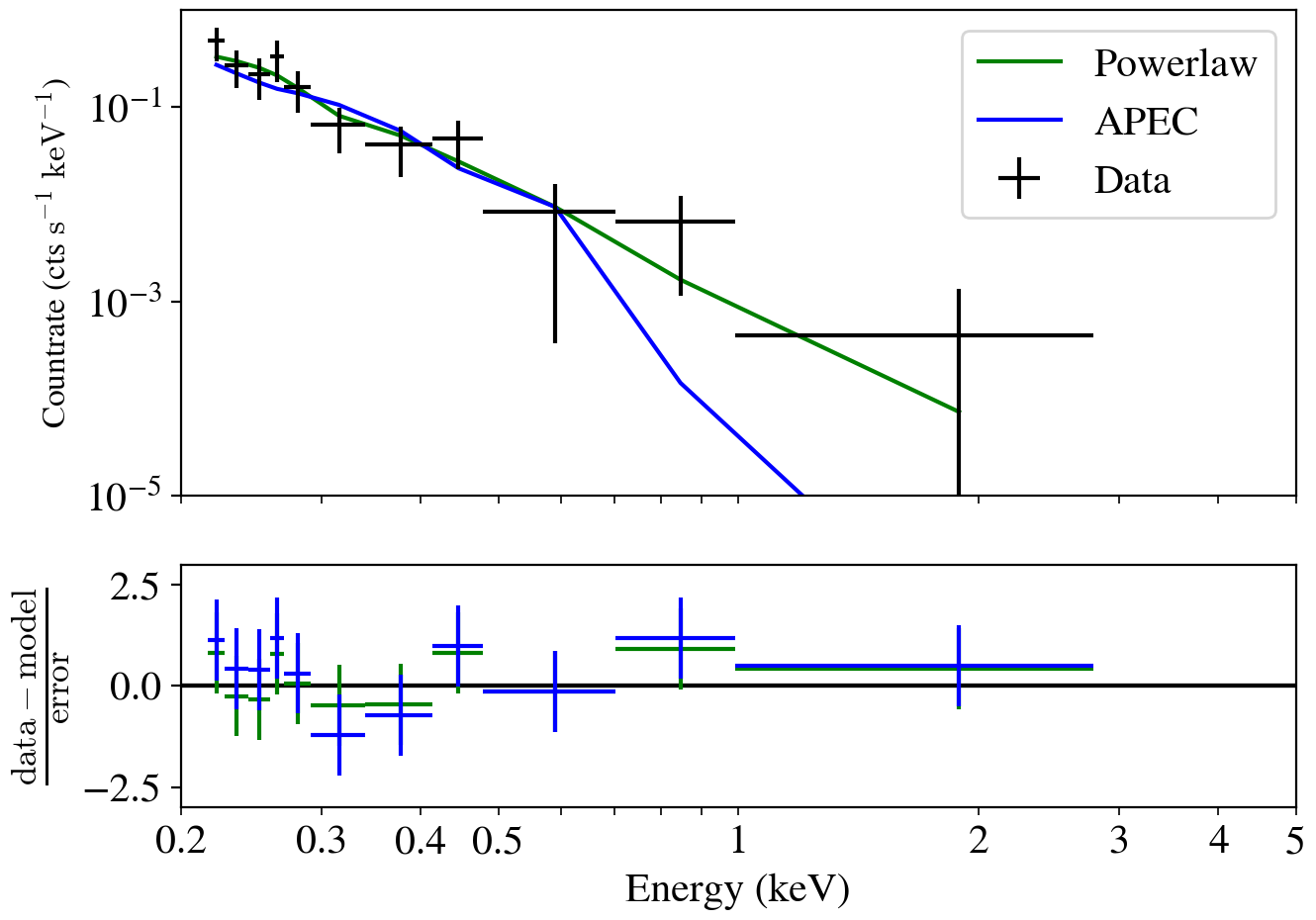}
    \caption{\textit{Upper panel:} eRASS:5 spectrum of 61\,Vir and power law fit (green) or 1T-APEC fit (blue). \textit{Lower panel:} residuals.
    }
    \label{fig:Spec61Vir}
\end{figure}

\paragraph{61 Vir}
With an eRASS count rate very close to the limit defined used to discard stars due to optical loading as shown in Fig.~\ref{fig:OpticalLoadingCut}, we suspect the G7-dwarf 61\,Vir to be likely affected by optical loading. To examine this, we fit the X-ray spectrum of the star with a power law (green) and compare it to a fit with a single-temperature APEC model (blue), as depicted in Fig.~\ref{fig:Spec61Vir}. We find that the spectrum is better described by a power law with a slope of $\alpha=6.0\pm0.8$. This is consistent with the value of $\alpha=6.1\pm0.1$ found by \citet{Robrade_2026} for detections caused by optical loading. The eRASS observations of the star are thus discarded.

\begin{figure}[h!]
    \centering
    \includegraphics[width=0.48\textwidth]{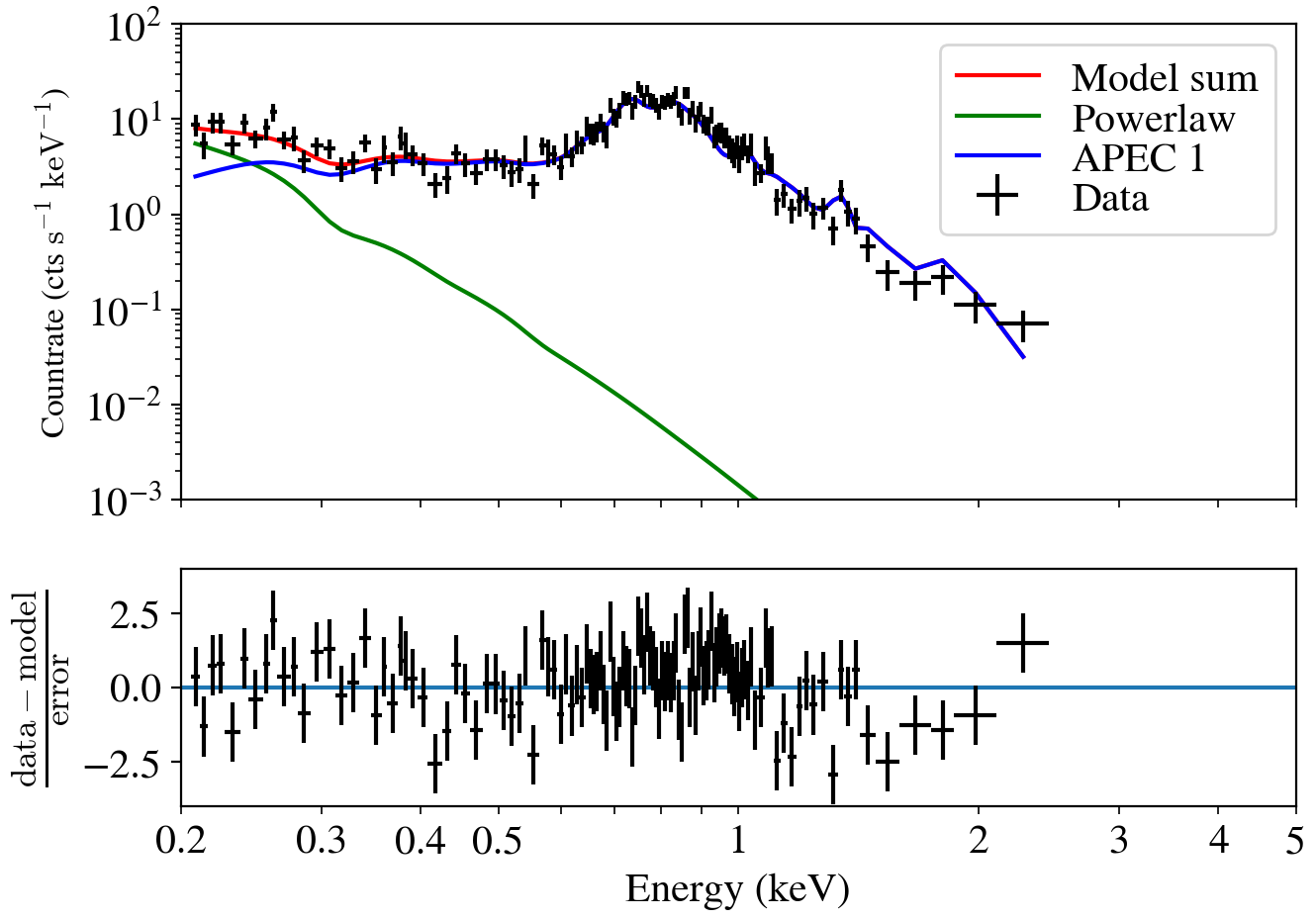}
    \caption{\textit{Upper panel:} eRASS4 spectrum of $\pi^3$\,Ori and model fit with power law and 1T-APEC components. \textit{Lower panel:} residuals.
    }
    \label{fig:Specpi3Ori}
\end{figure}

\begin{figure}[h!]
    \centering
    \includegraphics[width=0.48\textwidth]{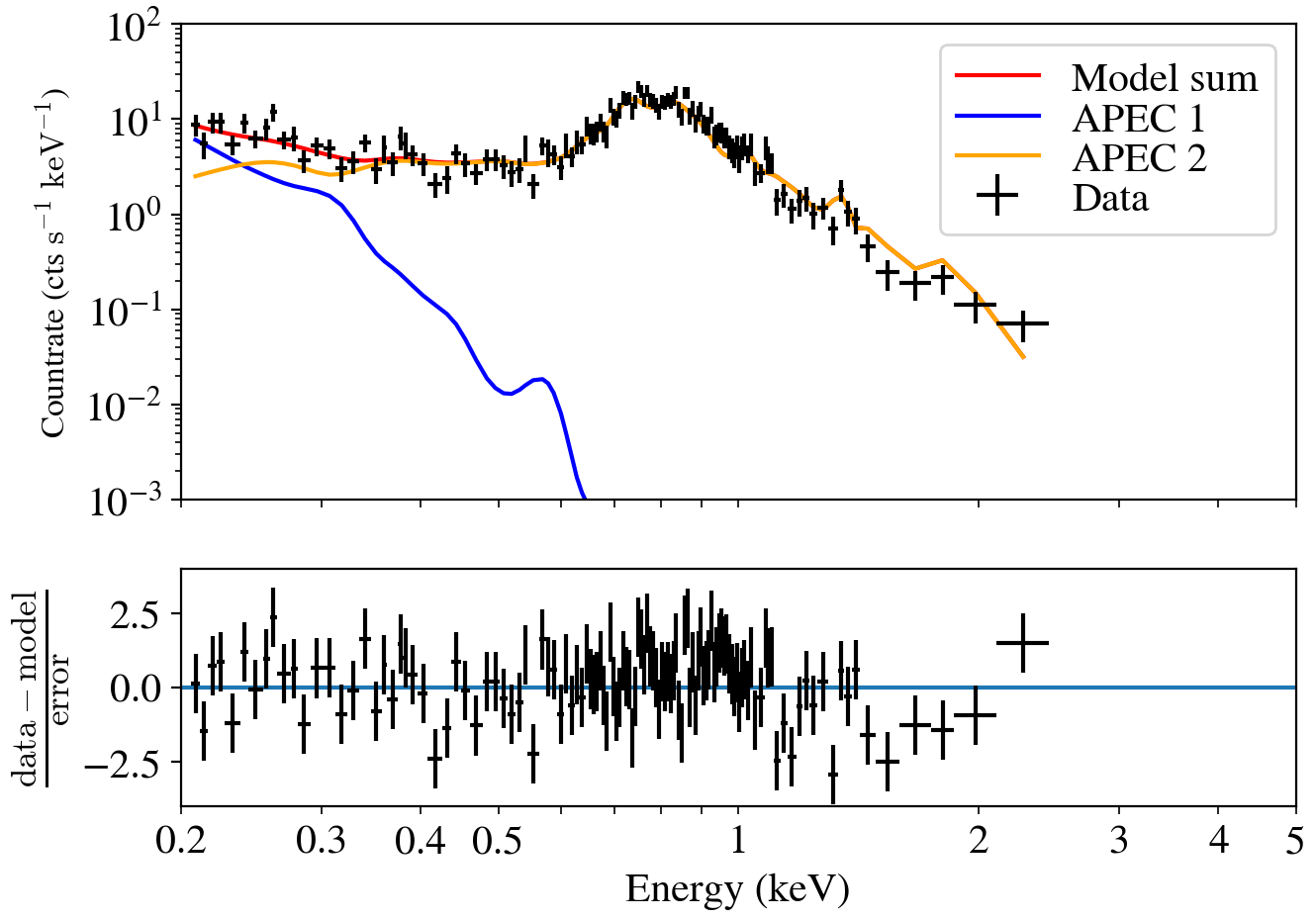}
    \caption{Same as Fig.~\ref{fig:Specpi3Ori}, but with 2T-APEC model instead.
    }
    \label{fig:Specpi3Ori_2TAPEC}
\end{figure}

\paragraph{$\pi^3$ Ori}
The F6-dwarf $\pi^3$\,Ori was observed in eRASS1--4 and is in the ``uncertain range'' regarding optical loading shown in Fig.\ref{fig:OpticalLoadingCut}. In an effort to retrieve spectral information from the star without the effects of optical loading, we model the X-ray spectra with a power law component. As is depicted in Fig.~\ref{fig:Specpi3Ori}, the spectrum is well-described ($\chi^2_{\mathrm{reduced}}\!=\!1.34$) by a model with a single-temperature APEC component and a power law with a slope of $\alpha=7.1\pm0.6$. However, the model is equally well-described ($\chi^2_{\mathrm{reduced}}\!=\!1.32$) with a two-temperature APEC model without a power law as presented in Fig.~\ref{fig:Specpi3Ori_2TAPEC}. In the latter case, the temperature of the cooler APEC component is very low at ${kT=0.04\,\mathrm{keV}}$. Due to the optical loading, we cannot constrain how much of the low-energy regime is attributed to low-temperature plasma.

\subsection{Data affected by optical loading}\label{app:NotInSample}
In Table~\ref{tab:DiscardedObservations}, we list the observations which we discarded due to contamination from optical loading as discussed in Sect.~\ref{subsect:opticalloading}.

\begin{table}	\caption{Observations which were discarded from the  analysis due to contamination from optical loading.}
\label{tab:DiscardedObservations}
\centering
\renewcommand*{\arraystretch}{1.25}
\begin{tabular}{c c c c}
\hline\hline
	X-ray source & Mission & Observation\\
	\hline
$\alpha$ Cen AB & eROSITA & eRASS1--5\\
HD 20794 & eROSITA & eRASS1--5\\
HD 219134 & \textit{XMM-Newton} & 0784920201\\
61 Vir & eROSITA & eRASS1--5\\
$\zeta$ Tuc & eROSITA & eRASS1--4\\
GJ 216 A & eROSITA & eRASS1--4\\
$\gamma$ Pav & eROSITA & eRASS1--4\\
HD 102365 AB & eROSITA & eRASS1--5\\ \hline
\end{tabular}
\end{table}

\section{Unresolved companions}\label{app:UnresolvedCompanions}

In Table~\ref{tab:UnresolvedCompanions}, we present the companions of FGK\,10pc stars which are unresolved within X-ray observations (see Sect.~\ref{subsect:XrayMultiples}). For our analysis, we determine the radii of these companion stars. If \textit{Gaia} photometry is available, we use isochrones from the \textsc{parsec v2.0} database \citep{Nguyen_2022, Nguyen_2025} where possible. The procedure was described by \citet{Bennedik_2026}. We adopt the metallicities from the parent star listed in Table~\ref{tab:FundamentalProperties} to constrain the \textsc{parsec} isochrones. 

If no \textit{Gaia} photometry is available, we instead adopt the radii listed in the table maintained by E. Mamajek, for which we adopt the SpT of the star from the catalogue by \citet{Reyle_2021}. However, a few stars do not have a SpT listed in this catalogue: For GJ\,570\,C we use the SpT of M3 as listed in the catalogue by \citet{Gonzalez_Payo_2026}. For 41\,Ara\,Bb we assume a SpT of M2.5 due to its estimated mass of ${0.41\,M_\sun}$ \citep{Tokovinin_2025}. For $\chi^1$\,Ori\,B we assume a SpT of M5 due to its mass of ${0.15\,M_\sun}$ \citep{Koenig_2002}. The SpT, radii and the method used for determining the radii are given in Table~\ref{tab:UnresolvedCompanions}.

\begin{table*}
\centering
\caption{Companion stars which are unresolved in X-ray observations. See text for descriptions of methods used to derive $R$.}
\label{tab:UnresolvedCompanions}
\begin{tabular}{l c c c c c c c}
\hline\hline
Star & Primary & SpT & $R$ & Method & \multicolumn{3}{c}{\textbf{Resolved from primary by:}}\\
 & & & ($R_\sun$) & & ROSAT & eROSITA & \textit{XMM-Newton} \\
\hline
61\,Cyg\,B & 61\,Cyg\,A & K7 & 0.57 & PARSEC & No & Undetected & Yes\\
70\,Oph\,B & 70\,Oph\,A & K4 & 0.68 & PARSEC & No & Undetected & No\\
GJ\,570\,B & GJ\,570\,A & M1 & 0.57 & PARSEC & No & Yes & Undetected\\
GJ\,570\,C & GJ\,570\,A & M3 & 0.57 & Mamajek & No & Yes & Undetected\\
$\eta$\,Cas\,B & $\eta$\,Cas\,A & K7 & 0.56 & PARSEC & No & Undetected & Undetected\\
36\,Oph\,B & 36\,Oph\,A & K1 & 0.72 & PARSEC & No & No & Undetected\\
GJ\,783\,B & GJ\,783\,A & M3.5 & 0.30 & Mamajek & No & Undetected & No\\
$\xi$\,Boo\,B & $\xi$\,Boo\,A & K5 & 0.62 & PARSEC & No & Undetected & Undetected\\
GJ\,105\,C & GJ\,105\,A & K7 & 0.12 & Mamajek & No & Undetected & Undetected\\
GJ\,53\,B & GJ\,53\,A & M4 & 0.27 & Mamajek & Undetected & Undetected & No\\
$\chi$\,Dra\,B & $\chi$\,Dra\,A & K0 & 0.81 & PARSEC & No & Undetected & Undetected\\
GJ\,66\,B & GJ\,66\,A & K2 & 0.72 & PARSEC & No & No & Undetected\\
$\chi^1$\,Ori\,B & $\chi^1$\,Ori\,A & M5 & 0.20 & Mamajek & No & No & Undetected\\
$\xi$\,UMa\,Ab & $\xi$\,UMa\,Aa & G2 & 0.36 & Mamajek & No & No & Undetected\\
$\xi$\,UMa\,Ba & $\xi$\,UMa\,Aa & K7 & 0.88 & PARSEC & No & No & Undetected\\
$\xi$\,UMa\,Bb & $\xi$\,UMa\,Aa & K7 & 0.77 & Mamajek & No & No & Undetected\\
41\,Ara\,Ba & 41\,Ara\,A & K9 & 0.51 & PARSEC & No & No & Undetected\\
41\,Ara\,Bb & 41\,Ara\,A & M2.5 & 0.42 & Mamajek & No & No & Undetected\\
HD\,102365\,B & HD\,102365\,A & M4 & 0.21 & PARSEC & Undetected & No; optical loading & Undetected\\
$\alpha$\,Cen\,B & $\alpha$\,Cen\,A & K1 & 0.87 & PARSEC & No & No; optical loading & No\\
GJ\,667\,B & GJ\,667\,A & K3 & 0.64 & PARSEC & No & No & Undetected\\
GJ\,667\,C & GJ\,667\,A & K5 & 0.54 & PARSEC & No & No & Undetected\\
\hline
\end{tabular}
\end{table*}

\section{Catalogue tables}
Here we provide the tables for the FGK\,10pc sample as described in Sect.~\ref{subsect:Catalogues}. Table~\ref{tab:FundamentalProperties} presents the fundamental properties and X-ray parameters. Individual observations are given in Table~\ref{tab:IndividualObservations}.

\begin{table*}
	\caption{Fundamental properties and mean X-ray parameters for the FGK\,10pc sample. ROSAT upper limits are marked with an asterisk.}
\label{tab:FundamentalProperties}
\centering
\renewcommand*{\arraystretch}{1.25}
\begin{tabular}{c c c c c c c c c c c c}
\hline\hline
	Name & SpT & Plx & [Fe/H] & ref & ${\rm log}\,L_{\rm bol}$ & $R$ & ${\rm log}\,L_{\rm X}$ & ${\rm log}\,F_{\!\rm X}$ & $kT$ & Multiplicity & X-ray source\\
	 &  & (mas) & (dex) &  & ($L_\sun$) & ($R_\sun$) & ($\rm{erg\,s^{-1}}$) & $\rm{(erg\,cm^{-2}\,s^{-1})}$ & (keV) &  & \\
	\hline
$\epsilon$ Eri & K2V & 310.58 & -0.08 & a & -0.48 & 0.75 & 28.35 & 5.82 & 0.35 & 1 & \\
61 Cyg A & K5V & 285.99 & -0.13 & a & -0.85 & 0.64 & 27.20 & 4.81 & 0.25 & 2 & \\
61 Cyg B & K7V & 286.01 & -0.21 & a & -1.06 & 0.57 & 27.11 & 4.81 & 0.27 & 2 & \\
$\epsilon$ Ind A & K5V & 274.84 & -0.13 & a & -0.68 & 0.68 & 27.47 & 5.02 & 0.24 & 3 & \\
$\tau$ Cet & G8.5V & 273.81 & -0.51 & a & -0.34 & 0.73 & 26.67 & 4.16 & 0.09 & 1 & \\
\vdots & \vdots & \vdots & \vdots & \vdots & \vdots & \vdots & \vdots & \vdots & \vdots & \vdots & \vdots\\
\hline
\end{tabular}
\tablefoot{This table is available in its entirety at the CDS. 
\\
\textbf{References.} 
a~\citet{Soubiran_2022};
b~\citet{Sousa_2008};
c~\citet{Marfil_2021};
d~\citet{Gaspar_2016}
}
\end{table*}

\begin{table*}
	\caption{X-ray parameters from individual observations. All \textit{XMM-Newton} observations are using EPIC/pn. For ROSAT 2RXP detections at large off-axis angles without uncertainties for the flux ($f_{\rm X}$), the uncertainties for ${\rm log}\,L_{\rm X}$ are computed only from parallax uncertainties.}
\label{tab:IndividualObservations}
\centering
\renewcommand*{\arraystretch}{1.25}
\begin{tabular}{c c c c c c c c c}
\hline\hline
	X-ray source & Mission & Observation & Mode & Epoch & Analysis & ${\rm log}\,f_{\rm X}$ & ${\rm log}\,L_{\rm X}$  & $\overline{kT}$\\
	 &  &  &  & (MJD) &  & ($\rm{erg\,cm^{-2}\,s^{-1}}$) & ($\rm{erg\,s^{-1}}$) & (keV)\\
	\hline
$\tau$ Cet & ROSAT & 2RXS &  & 48165.0 & Count rate & $-12.37^{0.08}_{0.10}$ & $26.83^{0.08}_{0.10}$ & \\
$\tau$ Cet & XMM-Newton & 0670380501 & Thick, Q & 55916.080 & 1T-APEC & $-12.79^{0.20}_{0.20}$ & $26.41^{0.20}_{0.20}$ & $0.09^{0.01}_{0.01}$\\
HD 88230 & ROSAT & 2RXS &  & 48165.0 & Count rate & $-11.89^{0.04}_{0.05}$ & $27.56^{0.04}_{0.05}$ & \\
HD 88230 & ROSAT & 2RXP &  & 48387.0 & Count rate & $-12.03$ & $27.43^{0.01}_{0.01}$ & \\
GJ 166 A & eROSITA & eRASS:5 &  & 59232.5 & 1T-APEC & $-12.31^{0.04}_{0.04}$ & $27.16^{0.04}_{0.04}$ & $0.27^{0.02}_{0.01}$\\
\vdots & \vdots & \vdots & \vdots & \vdots & \vdots & \vdots & \vdots & \vdots\\
\hline
\end{tabular}
\tablefoot{This table is available in its entirety at the CDS. }
\end{table*}

\section{Properties of Maunder minimum candidates}
In Table~\ref{tab:MMdiscussion}, we provide the fundamental properties of the Maunder minimum star HD\,166620 along with the three potential Maunder minimum candidates within the FGK\,10pc sample for our discussion in Sect.~\ref{subsubsect:MMstars}.

\clearpage
\begin{table*}
\centering
\caption{Properties of stars in the lower BKC area. Metallicities from \citet{Soubiran_2022}.}
\label{tab:MMdiscussion}
\begin{tabular}{l c c c c c c c c}
\hline\hline
Star & SpT & [Fe/H] & $P_\mathrm{rot}$ & ref & age & ref & $S_{\!{\rm HK}}$ & ref\\
 & & (dex) & (d) & & (Gyr) &  & median & \\
\hline
HD\,166620 & K2 & $-0.18\pm0.02$  & $45$ & (1) & 12.4 & (2) & 0.170 & (3)\\
$\tau$\,Cet & G8.5 & $-0.51\pm0.01$  & $46\pm4$ & (4) & $9.0\pm1.0$ & (5) & 0.167 & (6)\\
HD\,20794 & G8 & $-0.39\pm0.01$  & $\sim\!\!39$ & (7) & $\sim\!\!9$ & (8) & 0.168 & (7)\\
$\gamma$\,Pav & F9 & $-0.70\pm0.02$  & \multicolumn{2}{c}{Unknown} & $7.25 \pm 0.07$ & (9) & 0.265 & (10)\\
\hline
\end{tabular}
\tablebib{
(1)~\citet{Luhn_2022};
(2)~\citet{Brewer_2016};
(3)~\citet{Baum_2022}~(data 2004 onward);
(4)~\citet{Korolik_2023};
(5)~\citet{Tang_2011};
(6)~\citet{Isaacson_2024};
(7)~\citet{Nari_2025};
(8)~\citet{Medhi_2026};
(9)~\citet{Mosser_2008};
(10)~\citet{Saikia_2018}~(mean $S_{\!{\rm HK}}$)
}
\end{table*}

\end{appendix} 

\end{document}